\documentclass[12pt,twoside,english]{emulateapj}
\usepackage{mathptmx}
\usepackage[T1]{fontenc}
\usepackage{xcolor}
\usepackage{amsbsy}
\usepackage{amstext}
\usepackage{amssymb}
\usepackage{graphicx}
\PassOptionsToPackage{normalem}{ulem}
\usepackage{ulem}

\makeatletter

\providecommand{\tabularnewline}{\\}
\newcommand{\lyxdot}{.}

\providecolor{lyxadded}{rgb}{0,0,1}
\providecolor{lyxdeleted}{rgb}{0,0,1}

\usepackage{bm}

\AtBeginDocument{
  
}

\makeatother

\usepackage{babel}
\begin{document}
\global\long\def\Mbh{M_{\bullet}}
\global\long\def\ms{m_{\star}}
\global\long\def\Ns{N_{\star}}
\global\long\def\Ms{M_{\star}}
\global\long\def\Rs{R_{\star}}
 \global\long\def\tvRR{T_{\mathrm{vRR}}}
\global\long\def\tsRR{T_{\mathrm{sRR}}}
\global\long\def\xRR{x_{RR}}
\global\long\def\Rlso{r_{\mathrm{lso}}}
\global\long\def\Plso{P_{\mathrm{lso}}}
\global\long\def\Nlso{N_{\mathrm{lso}}}
\global\long\def\Mo{M_{\odot}}
\global\long\def\Mdot{\dot{m}}

\slugcomment{Draft version 4.0 \today}

\title{The torquing of circumnuclear accretion disks by stars\\
and the evolution of massive black holes}

\shorttitle{Torquing of circumnuclear accretion disks by stars and the evolution of massive black holes}

\author{Michal Bregman and Tal Alexander}

\shortauthors{Bregman \& Alexander}

\affil{Dept. of Particle Physics and Astrophysics, Faculty of Physics, Weizmann
Institute of Science, P.O. Box 26, Rehovot 76100, Israel}
\begin{abstract}
An accreting massive black hole (MBH) in a galactic nucleus is surrounded
by a dense stellar cluster. We analyze and simulate numerically the
evolution of a thin accretion disk due to its internal viscous torques,
due to the frame-dragging torques of a spinning MBH (the Bardeen-Petterson
effect) and due to the orbit-averaged gravitational torques by the
stars (Resonant Relaxation). We show that the evolution of the MBH
mass accretion rate, the MBH spin growth rate, and the covering fraction
of the disk relative to the central ionizing continuum source, are
all strongly coupled to the stochastic fluctuations of the stellar
potential via the warps that the stellar torques excite in the disk.
These lead to fluctuations by factors of up to a few in these quantities
over a wide range of timescales, with most of the power on timescales
$\gtrsim(\Mbh/M_{d})P(R_{d})$, where $\Mbh$ and $M_{d}$ are the
masses of the MBH and disk, and $P$ is the orbital period at the
disk's mass-weighted mean radius $R_{d}$. The response of the disk
is stronger the lighter it is and the more centrally concentrated
the stellar cusp. As proof of concept, we simulate the evolution of
the low-mass maser disk in NGC~4258, and show that its observed $O(10^{\circ})$
warp can be driven by the stellar torques. We also show that the frame-dragging
of a massive AGN disk couples the stochastic stellar torques to the
MBH spin and can excite a jitter of a few degrees in its direction
relative to that of the disk's outer regions. 
\end{abstract}


\section{Introduction}

\label{s:introduction}

Massive black holes (MBH) gain most of their mass, and presumably
a substantial fraction of their spin, in the course of luminous accretion
\citep{sol82,yuq+02}. This implies accretion via a radiatively efficient,
geometrically thin and optically thick accretion disk. As the MBH
mass grows, and in the absence of strong external perturbations (e.g.
galactic mergers), the stellar population around it is expected to
settle into a centrally concentrated cusp, $n_{\star}\propto r^{-\gamma}$
with $0.5\lesssim\gamma\lesssim2.5$, whether adiabatically (e.g.
\citealt{you80}), or by two-body relaxation (\citealt{bah+77,ale+09}).\texttt{\textbf{ }}This
stellar cluster, which spatially coexists with the disk and extends
farther out to a substantial fraction of the MBH's radius of dynamical
influence \citep[Fig. 12]{mer06}, can affect the disk both directly,
via hydrodynamical interactions \citep[e.g.][]{sye+91,art+93,vil+02},
and indirectly, by purely gravitational interactions \citep[hereafter BA09]{bre+09},
and thereby possibly influence the cosmic evolution of MBHs. As we
argue below (Section \ref{ss:OOM}), unless the disk is very compact,
it is the gravitational interactions that exert the dominant effect
by deforming the disk's geometry.

The best known example of a deformed thin Keplerian accretion disk
is that of the circumnuclear maser disk of active galaxy NGC~4258,
which displays an $O(10^{\circ})$ warp on the $0.1$ pc scale. In
a previous preliminary test of concept study (BA09), we analyzed the
physical scales in the problem, and argued that the observed warp
is consistent with resonant torquing of the disk by random fluctuations
in the stellar distribution around the MBH. Here we extend and generalize
that study by simulating numerically the evolution of a thin accretion
disk due to its internal viscous torques, due to the frame-dragging
torques of a spinning MBH and the stellar orbit-averaged gravitational
torques. We revisit the NGC~4258 maser disk, and further explore
the implications of stellar torquing for an accretion disk's geometry,
mass accretion rate and covering factor, and for the spin evolution
of MBHs.

The paper is organized as follows. In the remainder of the introduction
we describe a simplified model for a galactic nucleus harboring an
accreting MBH, and scale it to the NGC~4258 system (Section \ref{ss:NGC4258}).
We then briefly review Resonant Relaxation (RR, Section \ref{ss:RR})
and derive some relevant order of magnitude estimates (Section \ref{ss:OOM}).
In Section \ref{s:calculations} we describe how the MBH / disk /
stellar cluster system is modeled and simulated numerically (a brief
summary of the numerical scheme is presented in appendix \ref{a:discrete}).
We present the results in Section \ref{s:results} and discuss and
summarize them in Section \ref{s:discussion}.

\subsection{The NGC~4258 system}

\label{ss:NGC4258}

Maser-emitting nuclear accretion disks are unique, clean probes of
the environment near MBHs. The first discovered and best-studied maser
disk, NGC\,4258, is also the thinnest and most Keplerian (to better
than $1\%$; \citealt{mal02}) nuclear disk yet discovered. Radio
observations of $\mathrm{H_{2}O}$ maser emission from the edge-on
disk \citep{gre+95} reveal the disk morphology, gas velocities and
accelerations, and allow accurate measurements of the MBH mass ($\Mbh=3.7\times10^{7}\,\Mo$),
the distance to the host galaxy ($D=7.2\pm0.3$ Mpc) and the spatial
extent of the maser region ($R_{a}=0.13$ pc to $R_{b}=0.26$ pc)
\citep{her+96,her+99}. Together with optical and X-ray observations,
these also constrain the disk's accretion rate ($10^{-6}<\dot{M}<10^{-4}\,\Mo\,\mathrm{yr^{-1}}$)
\citep{neu+95}. The NGC\,4258 maser disk shows a clear $O(10^{\circ})$
warp \citep{her+96}. It is noteworthy that observations also indicate
possible warps in the maser disks of Circinus \citep{gre+03} and
NGC 3393 \citep{kon+08}.

Here the stellar cusp around the MBH is modeled as a single mass power-law
cusp of stars, $\Ns(<r)=N_{h}(r/r_{h})^{3-\gamma}$, which extends
inward from the radius of influence at $r_{h}=GM_{\bullet}/\sigma_{b}^{2}$,
where the velocity dispersion of the galactic bulge $\sigma_{b}$
can be estimated from the empirical $M_{\bullet}/\sigma$ relation
\citep{fer+00,geb+03,shi+03}, and contains $N_{h}=\mu_{h}\Mbh/\Ms$
stars of mass $\Ms$ each, where $\mu_{h}\sim O(1)$. We model NGC~4258
with $r_{h}=7$ pc, and assume $\mu_{h}=2$ (the formal value for
an MBH-less singular isothermal distribution), $\Ms=1\,\Mo$, and
unless stated otherwise, $\gamma=1.75$ (a dynamically relaxed cusp,
\citealt{bah+76}). For the purpose of simulating the torques on the
maser disk of NGC~4258, only stars in the radial range 0.01 pc to
$r_{h}$ are considered. The maser disk model is described in detail
in Section \ref{ss:disk}.

\subsection{Resonant relaxation}

\label{ss:RR}

Resonant relaxation (\citealt{rau+96,rau+98,hop+06a}) is a rapid
angular momentum relaxation mechanism that operates in potentials
with a high degree of approximate symmetry, which restricts orbital
evolution (e.g. fixed ellipses in a nearly Keplerian potential, or
fixed planar rosettes in a nearly spherical potential). In such potentials
the fixed, orbit-averaged, stellar mass distribution exerts a constant
residual torque on a test mass, which persists over a coherence time
$t_{0}$ as long as perturbations due to deviations from the perfect
symmetry remain small. The accumulated change in angular momentum
$\mathbf{J}$ over $t_{0}$ then becomes the {}``mean free path''
in $\mathbf{J}$-space for the noncoherent random walk phase on timescales
longer than $t_{0}$. When $t_{0}$ is long, the mean free path is
large and the random walk is rapid. The efficiency of RR is determined
by the nature of the physical process that perturbs the symmetry and
limits $t_{0}$ (e.g the Keplerian symmetry is perturbed far from
the MBH by the potential of the stars and near it by relativistic
precession). For circular orbits in a near spherical potential, such
as those of gas streams in an accretion disk, the RR torques can only
change the orbital orientation, but not the eccentricity, i.e. $J(r)=J_{c}(r)=\mathrm{const}$
\citep{gur+07}, where $J_{c}=\sqrt{G\Mbh r}$ is the maximal (circular)
angular momentum at radius $r$. On timescales $t<t_{0}$, the direction
of the angular momentum vector of a circular orbit of radius $r$
changes coherently due to the residual forces by stars on the same
scale, 
\begin{equation}
w(r)\equiv\left|\Delta J_{\perp}(r)\right|/J_{c}(r)=\beta_{\perp}\sqrt{\Ns(r)}(\Ms/\Mbh)t/P(r)\,,
\end{equation}
where $\Ns$ is the number of stars within $r$ and $P$ is the orbital
period. The numeric prefactor $\beta_{\perp}\sim O(1)$ may depend
somewhat on the parameters of the cusp, and can be determined from
$N$-body simulations, where it is measured to be $ $$\beta_{\perp}\gtrsim\sqrt{2}$
(\citealt{eil+09}; Kupi \& Alexander 2011, in prep.); $\beta_{\perp}=\sqrt{2}$
is adopted here. The {}``warp factor'' $w$ ($0\le w\le2$) corresponds
to a tilt in $\mathbf{J}$ by $\cos i=1-w^{2}/2$. 

In the limit where the disk mass $M_{d}$ is negligible compared to
the MBH mass $\Mbh$, the coherence time is set by the randomizing
effect of RR itself on the torquing stars ({}``self-quenching'').
In this case, the coherence time for the torque on a circular orbit
of radius $r$ by stars on the same scale is 
\begin{equation}
t_{\mathrm{sq}}(r)\sim A_{\mathrm{sq}}(\Mbh/\Ms)P(r)/\sqrt{\Ns(r)}\,,\label{e:tsq}
\end{equation}
 where $A_{\mathrm{sq}}$ is an $O(1)$ numeric prefactor. Preliminary
analysis of $N$-body simulations indicates that $A_{\mathrm{sq}}=1.0\pm0.1$
(Kupi \& Alexander 2011, in prep.); $A_{\mathrm{sq}}=1$ is adopted
here. The warp factor grows over the self-quenching time to $w(t_{\mathrm{sq}})=\beta_{\perp}A_{\mathrm{sq}}\simeq\sqrt{2}$,
that is, quenching occurs when the accumulated tilt in the orbital
direction grows to $i\simeq\pi/2$. An approximate correction for
the coherence time when the back reaction of the disk on the stars
cannot be neglected, is introduced in Section \ref{ss:OOM}.

\subsection{Order of magnitude estimates}

\label{ss:OOM}

We argue here, based on an approximate analysis of scales, that purely
gravitational interactions between the stars and the disk, rather
than hydrodynamic ones, typically dominate the torquing of the disk;
that around lower mass MBHs ($\lesssim10^{7}\,\Mo$ ) the disk warps
in response to the stellar torques, while around more massive MBHs
it changes its orientation as a nearly rigid body; and that stellar
torques are expected to warp the maser disk of NGC~4258 by $O(10^{\circ})$.

\paragraph{Direct stars / disk interactions}

Stars crossing the disk exert a torque on it by direct hydrodynamic
interaction. To assess whether this effect competes with the purely
gravitational RR torques, we estimate the torque exerted on the disk
in reaction to the ram pressure the disk exerts on the stars. The
residual torque by $\Ns$ stars of mass $M_{\star}$ and radius $R_{\star}$,
on randomly oriented orbits, in a volume of size $R$ enclosing a
disk of mass $M_{d}$, is
\begin{equation}
T_{\mathrm{ram}}\sim RF_{\mathrm{ram}}\sim R\sqrt{\Ns}\rho v^{2}\pi r_{X}^{2}\,,
\end{equation}
where $\rho$ is the disk density, $v^{2}\sim GM_{\bullet}/R$, $r_{X}=\max(\Rs,r_{B})$
is the effective cross-section radius of the star, where $r_{B}=2G\Ms/(v^{2}+c_{s}^{2})\simeq2(\Ms/\Mbh)R$
is the Bondi radius in the limit $c_{s}\lesssim O(10\,\mathrm{km\, s^{-1})}\ll v\sim O(10^{3}\,\mathrm{km\, s^{-1}})$,
which holds for a cold accretion disk close to the MBH. For $M_{\star}=1\, M_{\odot}$,
$R_{\star}=1\, R_{\odot}$, and for $M_{\bullet}=3.7\times10^{7}\, M_{\odot}$,
$R=0.26\,\mathrm{pc}$ and $H/R=0.002$ (the values for NGC~4258,
\citealt{her+05}), $r_{X}=\Rs$. The gas density can be estimated
by $\rho\sim M_{d}/(\pi R^{2}2H)$, where $H$ is the disk's exponential
scale height. We further assume that the cold thin disk is close to
its self-gravity stability limit (as implied by the non-smooth morphology
of many observed galactic maser disks, \citealt{mal02,bra+08}, and
is predicted to be the general case in AGN disks, \citealt{goo03}),
$M_{d}/M_{\bullet}\sim H/R$. The ram torque is then
\begin{equation}
T_{\mathrm{ram}}\sim R\sqrt{\Ns}\frac{M_{d}}{2\pi R^{2}H}\frac{GM_{\bullet}}{R}\pi R_{\star}^{2}=\frac{1}{2}\sqrt{\Ns}\frac{GM_{\bullet}^{2}}{R}\left(\frac{R_{\star}}{R}\right)^{2}\,,
\end{equation}
while the residual RR torque on the disk by the stars is
\begin{equation}
T_{\mathrm{RR}}\sim\sqrt{\Ns}\frac{GM_{\star}M_{d}}{R}\,.
\end{equation}
The RR torques affect the disk continuously over an orbital time,
whereas the ram torque is applied only twice per orbit, when the star
crosses the disk. The ratio between the durations the torques are
active is $\Delta t_{\mathrm{ram}}/\Delta t_{\mathrm{RR}}\sim4H/(2\pi R)$,
and therefore the ratio of accumulated changes in the disk angular
momentum per orbital time is 
\begin{equation}
\frac{\Delta J_{\mathrm{ram}}}{\Delta J_{\mathrm{RR}}}=\frac{T_{\mathrm{ram}}}{T_{\mathrm{RR}}}\frac{\Delta t_{\mathrm{ram}}}{\Delta t_{\mathrm{RR}}}=\frac{1}{\pi}\frac{M_{\bullet}}{M_{\star}}\left(\frac{R_{\star}}{R}\right)^{2}\,.
\end{equation}
For the parameters of NGC~4258, $\Delta J_{\mathrm{ram}}/\Delta J_{\mathrm{RR}}\sim10^{-7}$.
Even if AGN disks are limited by gravitational instability to radii
as small as $R\sim2000r_{g}$ ($r_{g}=G\Mbh/c^{2}$), as argued by
\citet{goo03}, then $\Delta J_{\mathrm{ram}}/\Delta J_{\mathrm{RR}}\sim0.02(\Mbh/10^{6}\,\Mo)^{-1}$,
and RR torquing dominates for all MBHs with mass $\Mbh\gtrsim\mathrm{few}\times10^{4}\,\Mo$.
The torquing of a thin disk by ram pressure is therefore negligible,
and the torquing effects of the stars on the disk are well approximated
by pure gravitational interactions.

\paragraph{Stellar torques vs. disk angular momentum transport}

The effect of the RR torques on the disk geometry depends on the ratio
between the stellar torques and the rate of angular momentum diffusion
in the disk. When the internal viscous torques are small, differential
torquing across the disk results in warps, which persist on the warp
diffusion timescale $t_{\mathrm{warp}}\sim(R/H)^{2}P/2\pi\alpha_{2}$,
where $\alpha_{2}\sim O(1)$ is the dimensionless viscosity parameter
associated with the vertical viscosity (Eq. \ref{e:a123}). The necessary
condition for the disk to be Keplerian, $\Ns\ll\Mbh/\Ms$, and the
condition for the warp to persist against diffusion, $t_{0}<t_{\mathrm{warp}}$,
provide together a soft upper limit to the mass range of MBHs where
disk warping is efficient, 
\begin{equation}
\Mbh<\Ms\left(\frac{R}{H}\right)^{4}\frac{f_{N}}{4\pi^{2}\alpha_{2}^{2}}\,,
\end{equation}
where $\Ns=f_{N}\Mbh/\Ms$ is the number of stars out of the $N_{h}\sim O(\Mbh/\Ms)$
in the MBH radius of influence that are close enough to the disk to
affect it appreciably. Conversely, when $t_{0}>t_{\mathrm{warp}}$,
the disk flattens out rapidly, and the stellar torquing results in
an overall evolution in the inclination of the disk as a nearly rigid
body. The upper limit for warping depends quite sensitively on the
assumed parameters, especially the disk's aspect ratio. For $R/H=250$,
$f_{N}=0.1$, $\Ms=1\,\Mo$ and $\alpha_{2}=1$, it is $\Mbh\sim O(10^{7}\,\Mo)$.

\paragraph{Residual stellar angular momentum}

A random density perturbation on the spatial scale $r$ has a Poisson
magnitude $\sqrt{\Ns(r)}$ and carries angular momentum of the order
$J_{N}\sim\sqrt{\Ns(r)}M_{\star}\sqrt{GM_{\bullet}r}$. This is the
maximal angular momentum that can be transferred to the disk from
that scale. The torque exerted by the stars on the disk will lead
to an equal and opposite torque by the disk on the stars. By the time
the disk has been torqued by $J_{N}$, the reaction force on the stars
will disperse the perturbation, and another, uncorrelated one, will
take its place. The back-reaction thus introduces a new coherence
timescale for the perturbation, $t_{\mathrm{react}}(r)$ such that
$\left|\Delta\mathbf{J}(t_{\mathrm{react}})\right|=J_{N}(r)$, which
may be shorter than the self-quenching coherence timescale, $t_{\mathrm{sq}}$
and thus limit the action of the torques on the disk. 

We estimate $t_{\mathrm{react}}$ here by approximating the disk,
which extends between $R_{\alpha}$ and $R_{\beta}$, as a thin ring
of mass $M_{d}$ and mass-weighted mean radius $R_{d}=2\pi M_{d}^{-1}\int_{R_{\alpha}}^{R_{\beta}}R^{2}\Sigma\mathrm{d}R$,
and evaluating the stellar torque on the disk in the limits of small
and large spatial scales. The lever arm on the gas disk is $R_{d}$.
When $r\gg R_{d}$, the magnitude of the force by the stars on the
disk is $\sqrt{\Ns(r)}GM_{\star}/r^{2}$, and so $\Delta J(t)/J_{c}(R_{d})=\beta_{\perp}A_{\mathrm{sq}}(R_{d}/r)^{1/2}t/t_{\mathrm{sq}}(r)$.
Conversely, when $r\ll R_{d}$, the force on the disk is $\sqrt{\Ns(r)}GM_{\star}/R_{d}^{2}$,
and so $\Delta J(t)/J_{c}(R_{d})=\beta_{\perp}A_{\mathrm{sq}}(r/R_{d})^{3/2}t/t_{\mathrm{sq}}(r)$.
In shorthand notation, 
\begin{equation}
\Delta J(t)/J_{c}(R_{d})=\beta_{\perp}A_{\mathrm{sq}}(r/R_{d})^{1/2-\Theta}t/t_{\mathrm{sq}}(r)\,,
\end{equation}
where $\Theta=\mathrm{sign}(r-R_{d})$. The back-reaction timescale
is then

\begin{equation}
t_{\mathrm{react}}(r)=\frac{\sqrt{\Ns(r)}}{\beta_{\perp}A_{\mathrm{sq}}}\frac{\Ms}{M_{d}}\left(\frac{r}{R_{d}}\right)^{\Theta}t_{\mathrm{sq}}(r)=\frac{1}{\beta_{\perp}}\frac{\Mbh}{M_{d}}\left(\frac{r}{R_{d}}\right)^{\Theta}P(r)\,,\label{e:treact}
\end{equation}
which increases monotonically with $r$. The coherence time is $t_{\mathrm{0}}(r,R_{d})=\min(t_{\mathrm{sq}},t_{\mathrm{react}})$.
Since $ $typically $t_{\mathrm{react}}(r,R_{d})/t_{\mathrm{sq}}(r,R_{d})=\sqrt{\Ns(r)}(\Ms/M_{d})/\sqrt{2}<1$
for $r\sim R_{d}$, where the stellar torques on the disk are most
effective, the back-reaction time is what sets the limit on the coherence
time. The change in the angular momentum of the disk over the back-reaction
time is

\begin{equation}
w_{\mathrm{react}}(r)=\sqrt{\Ns(r)}\frac{\Ms}{M_{d}}\sqrt{\frac{r}{R_{d}}}\,.
\end{equation}
For the NGC~4258 cusp parameters (Section \ref{ss:NGC4258}) and
a disk mass of $M_{d}\sim3000\, M_{\odot}$ (\citealt{mar08}; BA09)
with $r\sim R_{d}\simeq0.16$ pc, the predicted overall change in
disk orientation is $\sim27^{\circ}$, and the differential change
across the inner and outer edges of the masing region, the observed
warp angle $\omega$, is an order unity fraction of that (BA09). We
therefore anticipate that RR warping can lead to the $O(10^{\circ})$
warp that is observed in NGC~4258.

\section{Calculations}

\label{s:calculations}

\subsection{The evolution equation}

The equation governing the evolution of the angular momentum surface
density, $\mathbf{L}$, of a thin accretion disk under the influence
of internal and external torques, is given in the limits of a Keplerian
velocity field, no azimuthal modes and diffusive warp propagation
($\alpha_{1}>H/R$ ), by \citep{pap+83,pri92,ogi99} 
\begin{eqnarray}
\frac{\partial\mathbf{L}}{\partial t} & = & \frac{1}{R}\frac{\partial}{\partial R}\left[3R\frac{\partial}{\partial R}\left(\nu_{1}L\right)\boldsymbol{\ell}+\frac{1}{2}\nu_{2}RL\frac{\partial\boldsymbol{\ell}}{\partial R}\right]\nonumber \\
 &  & +\frac{1}{R}\frac{\partial}{\partial R}\left[\left(\nu_{2}R^{2}\left|\frac{\partial\boldsymbol{\ell}}{\partial R}\right|^{2}-\frac{3}{2}\nu_{1}\right)\mathbf{\mathbf{L}}\right]\nonumber \\
 &  & +\frac{1}{R}\frac{\partial}{\partial R}\left[\nu_{3}R\mathbf{\mathbf{L}}\times\frac{\partial\boldsymbol{\ell}}{\partial R}\right]\nonumber \\
 &  & +\mathbf{T}_{\mathrm{ext}}+\mathbf{T}_{\mathrm{src}}\,,\label{e:dLdt}
\end{eqnarray}
where $L=\Sigma\sqrt{GM_{\bullet}R}$, $\boldsymbol{\ell}=\mathbf{L}/L$
and $\Sigma$ is the disk's mass surface density. The disk is thus
effectively modeled as a set of concentric rigid thin annuli. The
first term on the right-hand side of Eq. (\ref{e:dLdt}) describes
the angular momentum that is carried to the central sink with the
inflowing mass, the second term the angular momentum that is advected
outward by the disk's internal viscous torques, which are expressed
by the azimuthal kinematic viscosity $\nu_{1}$ (responsible for the
mass inflow) and the vertical kinematic viscosity $\nu_{2}$ (responsible
for the unwarping of the disk). The third term describes the precessional
torque that accompanies large amplitude warps \citep{ogi99}. We assume
isotropic viscosities, $\nu_{n}=\widetilde{\alpha}_{n}c_{i}H$ ($n=1,2,3$),
where $c_{i}$ is the mid-plane isothermal sound speed, and use the
second order expansion of the three dimensionless viscosity parameters
\begin{equation}
\widetilde{\alpha}_{n}=A_{n}+B_{n}\psi^{2}+O(\psi^{4})\,,\qquad(n=1,2,3)\label{e:a123}
\end{equation}
 in terms of the local dimensionless local warp amplitude $\psi=R\left|\partial\boldsymbol{\ell}/\partial R\right|$
\citep{ogi99,lod+10}. The first order coefficients are
\begin{equation}
A_{1}=\alpha_{1}\,,\,\,\, A_{2}=\frac{2(1+7\alpha_{1}^{2})}{\alpha_{1}(4+\alpha_{1}^{2})}\,,\,\,\, A_{3}=\frac{3(1-2\alpha_{1}^{2})}{2(4+\alpha_{1}^{2})}\,.
\end{equation}
 We find that the 2nd order terms do not change the results significantly
in our simulations. The second order coefficients are listed for completeness
in appendix \ref{ss:ICBC}. For typical values $\alpha_{1}<1$, $\nu_{2}\gg\nu_{1}$,
that is, the disk unwarps faster than it flows into the MBH. 

$\mathbf{T}_{\mathrm{ext}}=\mathbf{T}_{\mathrm{BP}}+\mathbf{T}_{\mathrm{RR}}$
expresses the external torques per unit area applied to the disk,
which here are those due to relativistic frame-dragging and to the
stellar torques. \textcolor{black}{$\mathbf{T}_{\mathrm{src}}$ expresses
the torque per unit area due to a source term at the disk's outer
edge (Section \ref{ss:ICBC}).} The effect of frame dragging on the
disk, the Bardeen-Petterson (BP) effect, is given to lowest post-Newtonian
order by%
\footnote{The divergence of the BP torque toward the center is softened by substituting
$R\rightarrow\max(R,300r_{g})$.%
} \citep{bar+75} 
\begin{equation}
\mathbf{T}_{\mathrm{BP}}=\boldsymbol{\Omega}_{\mathrm{LT}}\times\mathbf{L}\,,\label{e:Tbp}
\end{equation}
where $\boldsymbol{\Omega}_{\mathrm{LT}}=(2G/c^{2}R^{3})\mathbf{J}_{\bullet}$
is the Lense-Thirring precession angular frequency and $\mathbf{J}_{\bullet}=\boldsymbol{\chi}GM_{\bullet}^{2}/c$
is the spin of the MBH, with $0\le|\boldsymbol{\chi}|\le1$ the dimensionless
spin parameter. The spin of the MBH evolves correspondingly by \citep{kin+05}
\begin{equation}
\mathrm{d}\mathbf{J}_{\bullet}/\mathrm{dt}=-2\pi\int_{R_{\alpha}}^{R_{\beta}}\mathrm{\mathbf{T}_{BP}}R\mathrm{d}R\,.\label{e:dJbhdt}
\end{equation}
Since the torque is perpendicular to the MBH spin, it changes only
its direction, not its magnitude. The BP torques work to align $\mathbf{L}$
and $\mathbf{J}_{\bullet}$. The frame dragging torques dominate over
the viscous torques out to the distance where the warp diffusion rate
becomes faster than the Lense-Thirring precession rate, $R_{\mathrm{BP}}=\sqrt{v_{2}(R_{\mathrm{BP}})/\Omega_{\mathrm{LT}}(R_{\mathrm{BP}})}$.
Typically $R_{\mathrm{BP}}\gg R_{\alpha}$.  The integrand in Eq.
(\ref{e:dJbhdt}) scales as $\Sigma(R)R^{^{-3/2}}$, so the torque
on the MBH is dominated by contributions from the innermost region
of the disk that remains non-aligned, and can be estimated by $\dot{J}_{\mathrm{BP}}\sim\pi R_{\mathrm{BP}}^{2}T_{\mathrm{BP}}(R_{\mathrm{BP}})=\pi\nu_{2}(R_{\mathrm{BP}})\Sigma(R_{\mathrm{BP}})\sqrt{G\Mbh R_{\mathrm{BP}}}$.
The timescale for the MBH spin to align itself with the disk is then
$t_{\parallel}\sim J_{\bullet}/\dot{J}_{\mathrm{BP}}$, which in steady-state
($\dot{M}\simeq3\pi\nu_{1}\Sigma$, $\nu_{2}\simeq\nu_{1}/2\alpha_{1}^{2}$)
can also expressed as 
\begin{equation}
t_{\parallel}=6\chi\alpha_{1}^{2}(\Mbh/\dot{M})\sqrt{r_{g}/R_{\mathrm{BP}}}\,.\label{e:talign}
\end{equation}
The torque on the MBH spin by direct accretion of matter, $\dot{J}_{\mathrm{acc}}\sim\dot{M}\sqrt{G\Mbh R_{\alpha}}$,
can be neglected relative to that by the BP effect, $\dot{J}_{\mathrm{BP}}$,
since $\dot{J}_{\mathrm{acc}}/\dot{J}_{\mathrm{BP}}\sim6\alpha_{1}^{2}\sqrt{R_{\alpha}/R_{\mathrm{BP}}}\ll1$.
Accretion can no longer be ignored on timescales long enough for it
to change the MBH mass appreciably, $t\gtrsim t_{E}=\eta c\sigma_{T}/4\pi Gm_{p}=\eta4.5\times10^{8}$
yr, where $t_{E}$ is the $e$-folding, or Salpeter, timescale for
growth by Eddington-limited accretion, $\sigma_{T}$ is the Thomson
cross-section, $m_{p}$ the proton mass and $\eta$ the radiative
efficiency of the accretion.

\subsection{The internal structure of the disk}

\label{ss:disk}

The internal structure of the disk needs to be specified to determine
the viscosity. In the $\alpha$-disk prescription, the mid-plane temperature
$T$ reflects the azimuthal kinetic viscosity of the disk via $\nu_{1}=\alpha_{1}c_{i}^{2}/\Omega_{K}$,
where $c_{i}^{2}=kT/\mu$ is the isothermal sound speed, $\mu$ is
the mean molecular weight and $\Omega_{K}=\sqrt{GM_{\bullet}/R^{3}}$
is the Keplerian angular frequency. The sound speed, in turn, determines
the disk's exponential scale height $H=c_{i}/\Omega_{K}$. 

The physical parameters of the masing region in the accretion disk
of NGC\,4258 can be inferred from the observed aspect ratio $H/R\sim0.002$,
which suggests a temperature of $T\sim600\,\mathrm{K}$ \citep{arg+07}
for a gas pressure supported disk (observations disfavor magnetic
support, see review by \citealt{lo05}), and from the fact that $\mathrm{H}_{2}\mathrm{O}$
maser emission is possible there. This requires the mid-plane molecular
hydrogen density $n_{\mathrm{H}_{2}}$ to be in the range $10^{7}\,\mathrm{cm^{-3}}$
to $10^{11}\,\mathrm{cm^{-3}}$ (all the hydrogen is assumed to be
molecular), the mid-plane gas temperature $T$ to be in the range
$300-400\,\mathrm{K}$ to $1000\,\mathrm{K}$, and the mid-plane pressure
$p/k$ to be in the range $10^{10}\,\mathrm{K\, cm^{-3}}$ to $10^{13}\,\mathrm{K\, cm^{-3}}$
\citep{mal02}. In addition, stability against fragmentation requires
that Toomre's criterion holds locally%
\footnote{Toomre's criterion can be recast as an approximate global condition,
$M(<R)/M_{\bullet}<c_{i}/v_{K}=H/R$, where $v_{K}=\sqrt{G\Mbh/R}$. %
}, $Q=c_{i}\Omega_{K}/\pi G\Sigma>1$ \citep{too64}, which further
constrains the disk mass and temperature.

It is unclear whether the accretion disk's internal viscosity alone
can provide enough heat to warm the molecular gas to masing temperatures.
\citet{neu+95} argue that heating by X-ray irradiation by the central
source is essential for creating the required conditions in the masing
region. Outside the masing region, the disk's properties are not usefully
constrained by current observations.

Here we make a choice of convenience to adopt the known solution%
\footnote{The steady state solution has $T\propto R^{-3/4}$, $\Sigma\propto R^{-3/4}$,
$c_{i}\propto R^{-3/8}$, $H\propto R^{9/8}$, $n_{H_{2}}\propto R^{-15/8}$,
$\nu{}_{1}\propto\nu_{2}\propto R^{+3/4}$, $p/k\propto R^{-21/8}$
and Toomre's $Q\propto R^{-9/8}$ .%
} of a gas pressure-dominated thin accretion disk, where the temperature
is completely determined by the internal viscosity (that is, no external
heating) and free-free (Kramer's law) opacity \citep{sha+73}. The
prefactor $\kappa_{a}$ of Kramer's law, $\kappa=\kappa_{a}(\rho/\rho_{a})(T/T_{a})^{-7/2}$,
where $\rho=\Sigma/(\sqrt{2\pi}H)$ is the total mid-plane mass density
(the subscript $a$ denotes values at $R_{a}$) is then adjusted to
be high enough to maintain the required temperature at the masing
region%
\footnote{The actual opacity law in a cool molecular disk presumably resembles
that of a proto-planetary disk, e.g. \citet{sem+03}, which does not
have a simple analytic form. %
}. We assume Solar abundances ($X=0.7057$, \citealt{arn96}) and that
all the hydrogen is molecular, which corresponds to a mean molecular
weight of $\mu=2.358m_{p}$. The adopted value $[H/R]_{a}=0.002$
fixes the surface density $\Sigma{}_{a}=\rho_{a}\sqrt{2\pi}[H/R]_{a}R_{a}$,
where $\rho_{a}=n_{\mathrm{H}_{2}}(R_{a})2m_{p}/X$. We find that
in order to have masing conditions between $R_{a}$ and $R_{b}$,
it is necessary to assume the highest temperature value at the inner
edge, $T_{a}=10^{3}\,\mathrm{K}$. The assumed density there, $n_{H_{2}}(R_{a})=3\times10^{8}\,\mathrm{cm^{-3}}$,
which is within the masing range and consistent with the limits on
$p_{a}/k=(\rho_{a}/\mu)T_{a}$, is chosen so the total disk mass is
$3\times10^{3}\,\Mo$. The mid-plane temperature is given by $\sigma_{\mathrm{SB}}T^{4}=(27/32)\Sigma^{2}\Omega_{K}^{2}\nu_{1}\kappa$
(e.g. \citealt{fra+02}, Eq. 5.76), which here yields 

\begin{equation}
T(R,t)=\left[\frac{27k_{B}^{1/2}}{\sqrt{2\pi}32\sigma_{\mathrm{SB}}}\frac{\alpha_{1}\kappa_{a}}{\mu{}^{1/2}}\frac{T_{a}^{7/2}}{\rho_{a}}\Sigma^{3}(R,t)\Omega_{K}^{2}(R)\right]^{1/7}\,.\label{e:TcR}
\end{equation}
For the values of $T_{a}$ and $\rho_{a}$ above, Eq. (\ref{e:TcR})
requires for self-consistency $\kappa_{a}=7.195\times10^{5}/\alpha_{1}\,\mathrm{cm^{2}\, g^{-1}}$.
This is an extremely high near/mid IR opacity ($\lambda_{\max}\sim3-10$
$\mu\mathrm{m}$ for $T\sim300$--$1000\,\mathrm{K}$), as compared,
for example, to the mean theoretical opacity of $\left\langle \kappa(2.2\,\mathrm{\mu m})\right\rangle =3800\pm700\,\mathrm{cm}^{2}\,\mathrm{g}^{-1}$$ $
used to model dusty star forming cores \citep{shi+11}. This is in
line with the conclusion that an external source of heating is required
to explain the gas temperature.

\subsection{Numerical implementation}

The finite difference method is used to integrate the evolution equations
(Eq. \ref{e:dLdt} with $G=c=M_{\bullet}=1$) for the angular momentum
surface densities $\{\mathbf{L}_{i}^{j}\}_{i=0}^{N+1}$ at time $t_{j}$,
at positions $\{R_{i}\}_{i=0}^{N+1}$ on a logarithmically spaced
grid between $R_{1}=R_{\alpha}$ at the innermost stable circular
orbit (ISCO) and $R_{N}=R_{\beta}\gg R_{1}$. The edge points $R_{0}$
and $R_{N+1}$ are used to enforce the boundary conditions. A brief
description of the numerical scheme \citep{pap+83,pri92} is given
in appendix \ref{a:discrete}.

\subsubsection{Initial conditions and boundary conditions}

\label{ss:ICBC}

The initial conditions are a flat disk with surface density $\Sigma=\Sigma_{a}(R/R_{a})^{-3/4}$
and normal $\boldsymbol{\ell}_{0}$, where $\Sigma$ is chosen so
as to satisfy the maser conditions between the masing region limits
$R_{a}$ and $R_{b}$. The MBH spin is initially along the $z$-axis.
We assume the boundary conditions $L(R_{1})=0$ and $\partial\boldsymbol{\ell}/\partial R|_{R_{1}}=0$
(central mass sink and no warp), and $\partial(\nu_{1}\mathbf{L})/\partial R|_{R_{2}}=0$
(no torque). These boundary conditions allow mass through the inner
boundary into the MBH. To prevent the disk from draining on the viscous
timescale, and from drifting away from the masing conditions much
before that, a source term is added at the outer edge. It is adjusted
every time-step $t_{j}$ to compensate for the mass lost from the
disk, $(\Delta M)^{j}=\sum_{i=1}^{N}(M_{i}^{j}-M_{i}^{j-1})$, where
$M_{i}^{j}=2\pi R_{i}\Delta R_{i}\Sigma_{i}^{j}$ is the mass in a
disk ring of width $\Delta R_{i}=R_{i}-R_{i-1}\ll R_{i}$. The angular
momentum density at the outermost grid point is then augmented by
an amount $(\Delta L_{N}){}^{j}\boldsymbol{\ell}_{0}$, such that
$\left|\mathbf{L}_{N}^{j}+(\Delta L_{N})^{j}\boldsymbol{\ell}_{0}\right|=\left[\Sigma_{N}^{j}+(\Delta\Sigma_{N})^{j}\right]\sqrt{R_{N}}$,
where $\left(\Delta\Sigma_{N}\right)^{j}=-X_{M}(\Delta M)^{j}/2\pi R_{N}\Delta R_{N}$
($X_{M}$ is an order unity stabilization factor, see below). The
magnitude of the angular momentum needed to compensate for the mass
loss is therefore 
\begin{eqnarray}
(\Delta L_{N})^{j} & = & -\mathbf{L}_{N}^{j}\cdot\boldsymbol{\ell}_{0}+\nonumber \\
 &  & \sqrt{(\mathbf{L}_{N}^{j}\cdot\boldsymbol{\ell}_{0})^{2}+[\Sigma_{N}^{j}+(\Delta\Sigma_{N})^{j}]^{2}R_{N}-(L_{N}^{j})^{2}}\,.\label{e:dLN}
\end{eqnarray}
The fluctuating factor $X_{M}=1\pm\epsilon$ stabilizes the disk against
the accumulation of numeric integration errors by allowing small over-
or under-corrections, as needed%
\footnote{This algorithm of stabilization by over-shooting is inspired by the
engineering method of control system hysteresis, used e.g. in thermostats. %
}. When $(\Delta M)^{j}<0$, $X_{M}$ is set to $1+\epsilon$, and
conversely, when $(\Delta M)^{j}>0$, it is set to $1-\epsilon$.
Experimentation indicates that $\epsilon=0.1$ is a good choice. 

We find that with the source term thus defined, a flat disk with the
inner boundary conditions $\nu_{1}\Sigma|_{R_{1}}=0$, rapidly converges
to a close approximation of its theoretically expected steady state
solution with the expected mass-loss rate of $\dot{M}\simeq3\pi\nu_{1}\Sigma$
(for $R\gg R_{\alpha}$). Likewise, the mass of a non-stationary disk
rapidly converges to a steady state value that is typically within
$O(10^{-3})$ of its initial mass, even as its geometry continuously
changes. The mean accretion rate over some period $t_{j_{1}}$ to
$t_{j2}$ can then be estimated by $\left\langle \dot{M}\right\rangle =(\sum_{j=j_{1}}^{j_{2}}X_{M}(\Delta M)^{j})/(t_{j2}-t_{j1})$.

This mass replenishment scheme represents an idealized case of a self-regulating
mass supply, which prevents both the draining and over-loading and
fragmentation of the disk. It should be emphasized that this assumption
is introduced here for convenience only, to stabilize the disk long
enough to collect robust statistics. Our statistical conclusions about
the dynamical response of the disk to the external stellar torques
do not depend strongly on the existence of such a mass source, as
long as the disk mass is in quasi-equilibrium for at least the relatively
short RR timescale.

\subsubsection{The stellar resonant relaxation torques}

\label{ss:RRcusp}

The RR torques due to the $O(10^{8}$) stars in the radius of influence
cannot be modeled directly. Instead, they are approximated here by
representing the stellar cusp by a small number of concentric spherical
shells delimited by the radii $\left\{ r_{k}\right\} _{k=0}^{N_{s}}$
(where $r_{0}=0$). The spacing between consecutive shell radii is
chosen to be large enough so that the residual forces are approximately
independent of each other, $r_{k+1}/r_{k}\ge2^{2/(3-\gamma)}$, which
corresponds to the requirement that $\sqrt{N_{\star}(<r_{k})}\ge2\sqrt{N_{\star}(<r_{k-1})}$
(BA09), and is broadly consistent with the rigorous derivation of
\citet{koc+11}. The maximal number of such logarithmically spaced
independent star shells in the simulations of $ $NGC~4258 that are
presented below is typically $N_{s}=5$. The vector RR torques from
each shell are represented by the gravitational field of a thin ring
of mass $M_{k}=\sqrt{N_{\star}(<r_{k})-N_{\star}(<r_{k-1})}\Ms$ with
radius $\bar{r}_{k}=(r_{k}+r_{k-1})/2$ and normal $\mathbf{n}_{k}(t)$,
which is interpolated smoothly in time by cubic spline from a sequence
of isotropic random normals $\{\mathbf{n}_{k}^{j}\}_{j=0}^{[t_{\mathrm{sim}}/t_{0}]+1}$
at times $t_{j}=jt_{0}(r_{k})$, where $t_{\mathrm{sim}}$ is the
duration of the simulation and $t_{0}(r)=\min(t_{\mathrm{sq}}(r),t_{\mathrm{react}}(r))$
(Eqs. \ref{e:tsq}, \ref{e:treact}) is the coherence time. This simulates
the random reorientation of the local residual force after a vector
RR timescale. Finally, the torque exerted by a star ring at $\bar{r}_{k}$
on a disk gas ring at $R_{i}$, which is directed along the line of
intersection of the two rings, is calculated numerically by integrating
over the gravitational force between all mass elements in the rings
\citep{nay05}, 

\begin{eqnarray}
\mathbf{T}_{\mathrm{RR}}(k & \rightarrow & i)=\frac{M_{i}M_{k}R_{i}\bar{r}_{k}}{(R_{i}^{2}+\bar{r}_{k}^{2})^{3/2}}\frac{\sin\beta}{4\pi^{2}}\times\\
 &  & \int_{0}^{2\pi}\mathrm{d}\phi_{1}\int_{0}^{2\pi}\mathrm{d}\phi_{2}\frac{\sin\phi_{1}\sin\phi_{2}}{[1-\delta\cos\lambda]^{3/2}}(\boldsymbol{\ell}_{i}\times\mathbf{n}_{k})\,,
\end{eqnarray}
where $M_{i}$ is the mass of the disk annulus, $\beta=\cos^{-1}(\boldsymbol{\ell}_{i}\cdot\mathbf{n}_{k})$,
$\delta=1-(\bar{r}_{k}^{2}-R_{i}^{2})/(\bar{r}_{k}^{2}+R_{i}^{2})$
and%
\footnote{The divergence of the torque when $\bar{r}_{k}\rightarrow R_{i}$
is smoothed by substituting $\delta\rightarrow\delta=1-\max\left[(\bar{r}_{k}^{2}-R_{i}^{2}),(\bar{r}_{k}-\bar{r}_{k-1})^{2},(R_{i}-R_{i-1})^{2}\right]/(\bar{r}_{k}^{2}+R_{i}^{2})$. %
} $\cos\lambda=\cos\beta\sin\phi_{1}\sin\phi_{2}+\cos\phi_{1}\cos\phi_{2}$.

\subsubsection{Validation of the code}

\label{ss:validation}

We verified that the numeric integration scheme conserves angular
momentum to floating point precision, and conserves mass to a fractional
precision of $O(10^{-10}\,\mathrm{yr^{-1}})$, by switching off the
source term and evolving a flat disk from an out-of-steady-state initial
surface mass density profile ($\Sigma\propto R^{-1}$). We also evolved
an $\Sigma\propto R^{-3/4}$ initial configuration with a source term
at $R_{2}$ and with boundary conditions $\nu_{1}\Sigma=0$ at $R_{1}$,
and verified it approximates the analytical solution $3\pi\nu_{1}\Sigma=\dot{M}(1-\sqrt{R_{1}/R})$
(e.g. \citealt{fra+02}) reasonably well (the match with the analytic
solution is better on a linear grid than on a logarithmic one; overall
the match on a logarithmic grid is at a level similar to that obtained
by \citealt{pri92}). We also reproduce qualitatively the response
of the disk to initial twists and to the BP torques that are presented
by \citet{pri92}.

\section{Results}

\label{s:results}

\begin{figure*}
\noindent \begin{centering}
\begin{tabular}{cc}
(a)\includegraphics[width=0.45\textwidth]{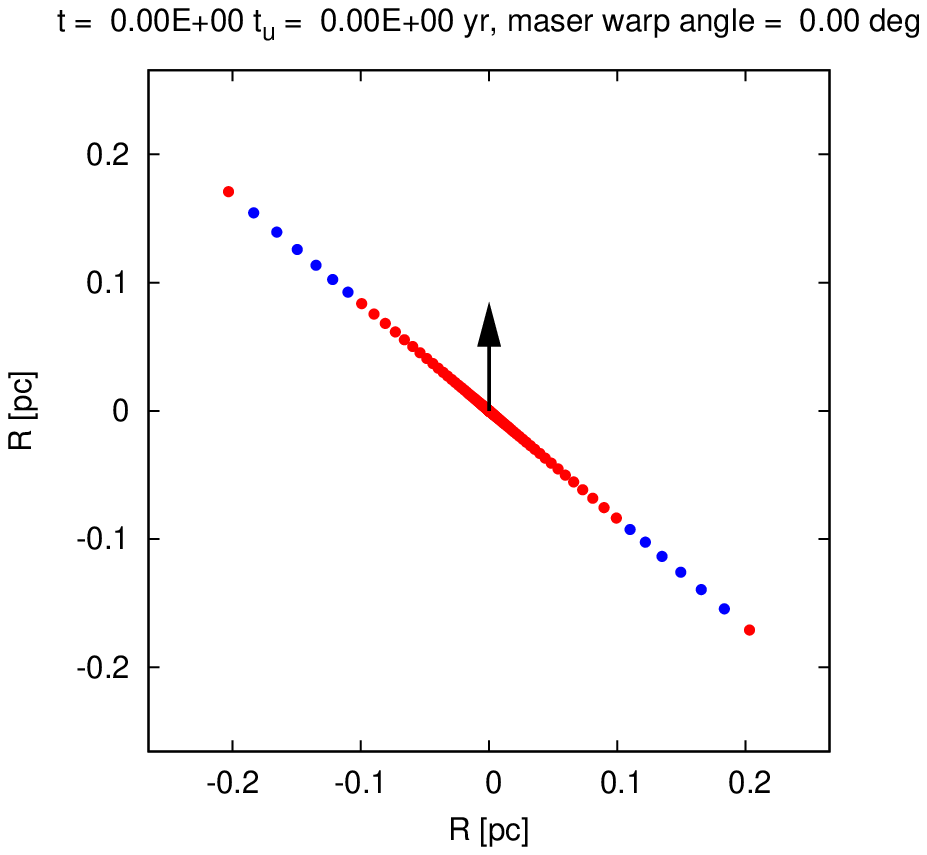} & (b)\includegraphics[width=0.45\textwidth]{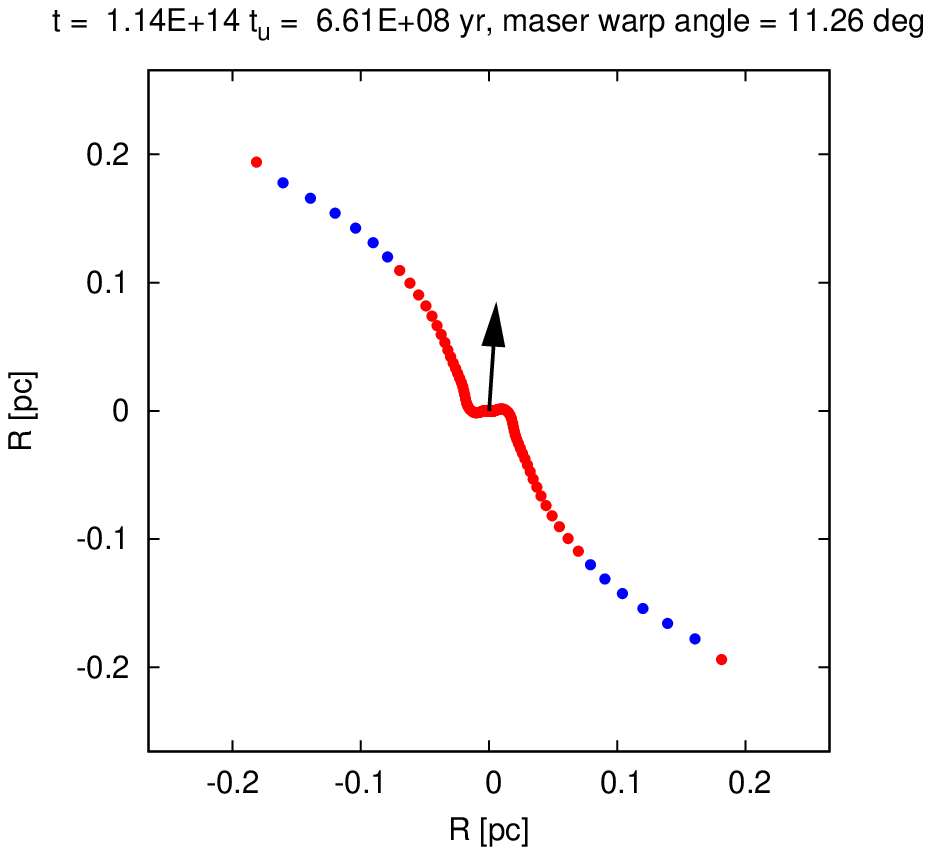}\tabularnewline
(c)\includegraphics[width=0.38\textwidth]{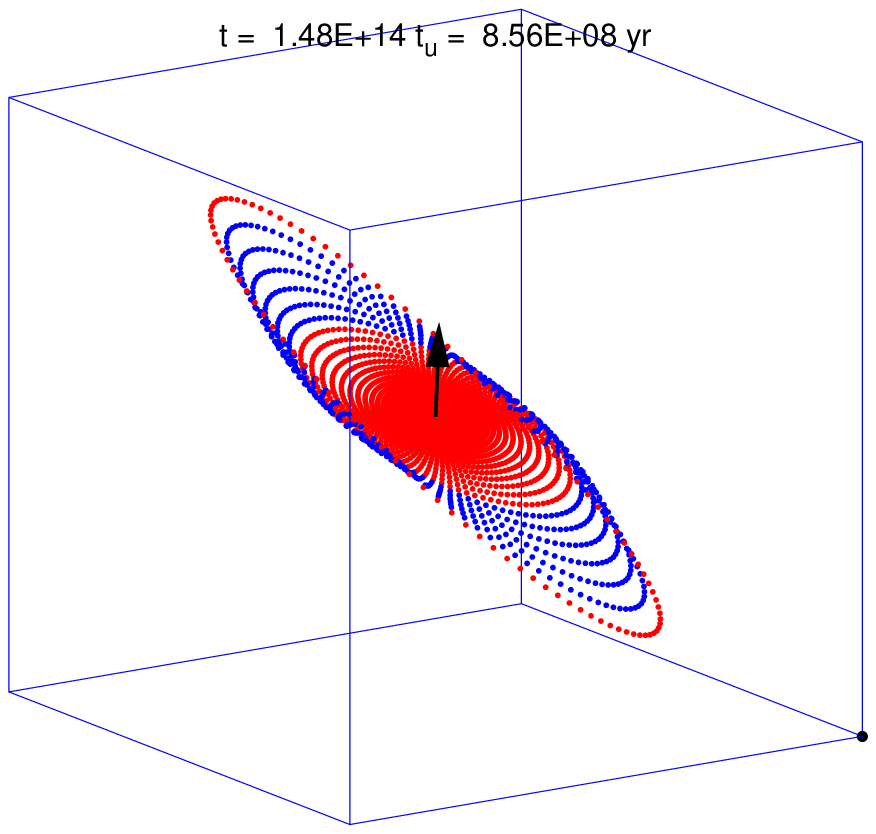} & (d)\includegraphics[width=0.48\textwidth]{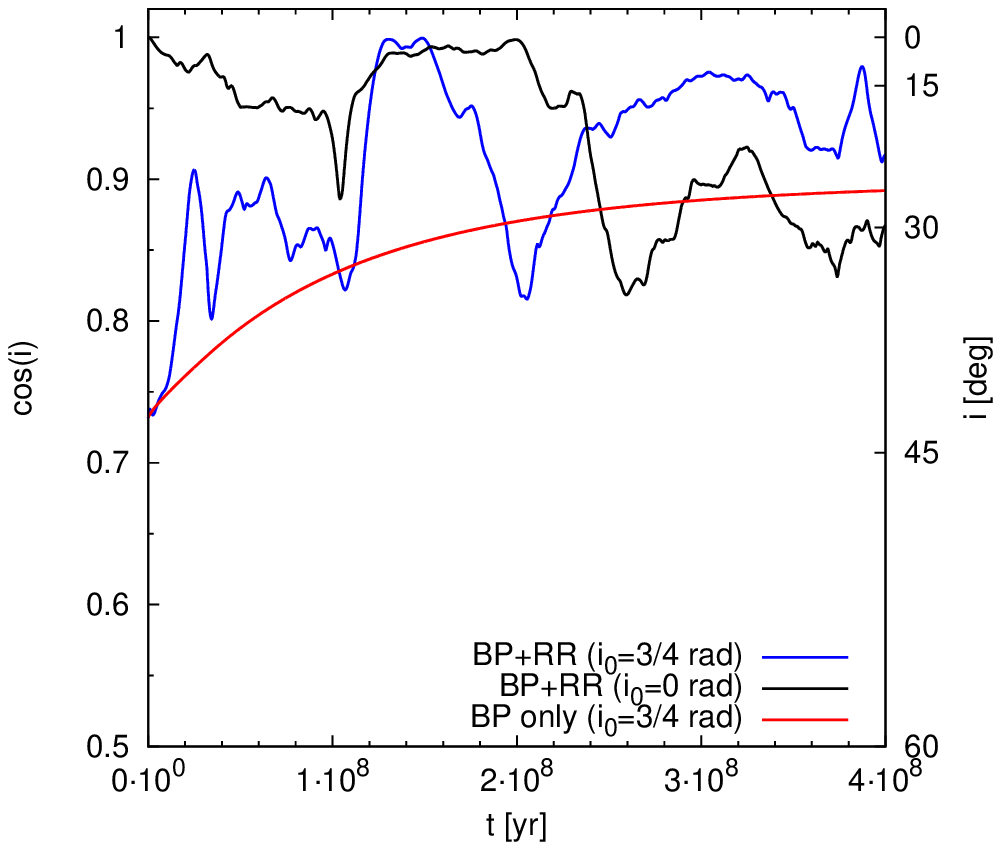}\tabularnewline
\end{tabular}
\par\end{centering}

\caption{\label{f:snaps} Top row and bottom left: Snapshots from a disk evolution
simulation of NGC\,4258 with $\alpha_{1}=0.25$, $\chi=1$, Kramer's
opacity law, and the initial conditions $\mathbf{J}_{\bullet}=\hat{z}$
and $i_{0}=3/4$ rad. The arrow denotes the MBH spin vector. The masing
region is marked in blue. Top: The intersection of the disk mid plane
with the $x-z$ plane (the simulated plane of sky). Bottom left: The
disk mid plane in 3D. Bottom right: The evolution of $\cos i$ in
the simulation shown in the snapshots, and for similar simulations
with $i_{0}=0$ rad, and with BP torques only ($\Ns=0$).}
\end{figure*}

We carried out a suite of disk evolution simulations, using a stellar
cusp model based on NGC\,4258 (Section \ref{ss:NGC4258}) on a logarithmic
grid with with $N=100$ points extending from $R_{1}=6r_{g}$ to $R_{N}=1.5\times10^{5}r_{g}$
($r_{g}\simeq5.5\times10^{12}\,\mathrm{cm}\simeq1.8\times10^{-6}\,\mathrm{pc}$).
The typical time-step increments in our simulations were in the range
$\Delta t_{j}\sim0.01$--$1$ yr. The initial disk configuration was
flat with a $\Sigma\propto R^{-3/4}$ surface mass density profile,
and the disk was tilted by an angle $i_{0}$ relative to the $z$-axis,
which coincided with the initial MBH spin, in those simulations with
$\chi\neq0$. In the simulations presented below, we explore the initial
conditions $i_{0}=0,3/4$ rad and $\chi=0,1$. At later times, we
generalize the definition of the tilt angle to $i=\cos^{-1}(\mathbf{J}_{\bullet}\cdot\mathbf{J}_{\mathrm{disk}}/J_{\bullet}J_{\mathrm{disk}})$,
where $\mathbf{J}_{\mathrm{disk}}$ is the total angular momentum
of the disk. To isolate the effect of the BP torques on the disk,
we ran some simulations with $N_{\star}=0$. In the presence of RR,
the disk never reaches a steady state, and the simulations are terminated
after they are observed to reach statistical stationarity (recall
that the mass supply rate is adjusted continuously to maintain constant
total mass, Section \ref{ss:ICBC}). Simulations without RR are stopped
when the relative rate of change in $\left\{ \mathbf{L}_{i}\right\} _{i=1}^{N}$
falls below some very small value.

Figure (\ref{f:snaps}) shows snapshots from a simulation of a misaligned
disk ($i_{0}=3/4$ rad) evolving under both BP and RR torques, and
the evolution of the inclination angle $i$ for this and two other
simulations, one with no RR and the other starting with an aligned
disk ($i_{0}=0$ rad). The evolution from the initial (and artificial)
misaligned configuration, where the disk is tilted relative to the
MBH spin all the way down to $R_{1}$, shows strong stochastic RR-dominated
evolution that is superposed on a slower frame dragging-dominated
secular evolution, until at later times the stochastic behavior dominates.

\subsection{Disk warping}

\label{ss:warp}

\begin{figure*}
\noindent \begin{raggedright}
\begin{tabular}{c}
$\!\!\!\!\!$\includegraphics[width=1\textwidth]{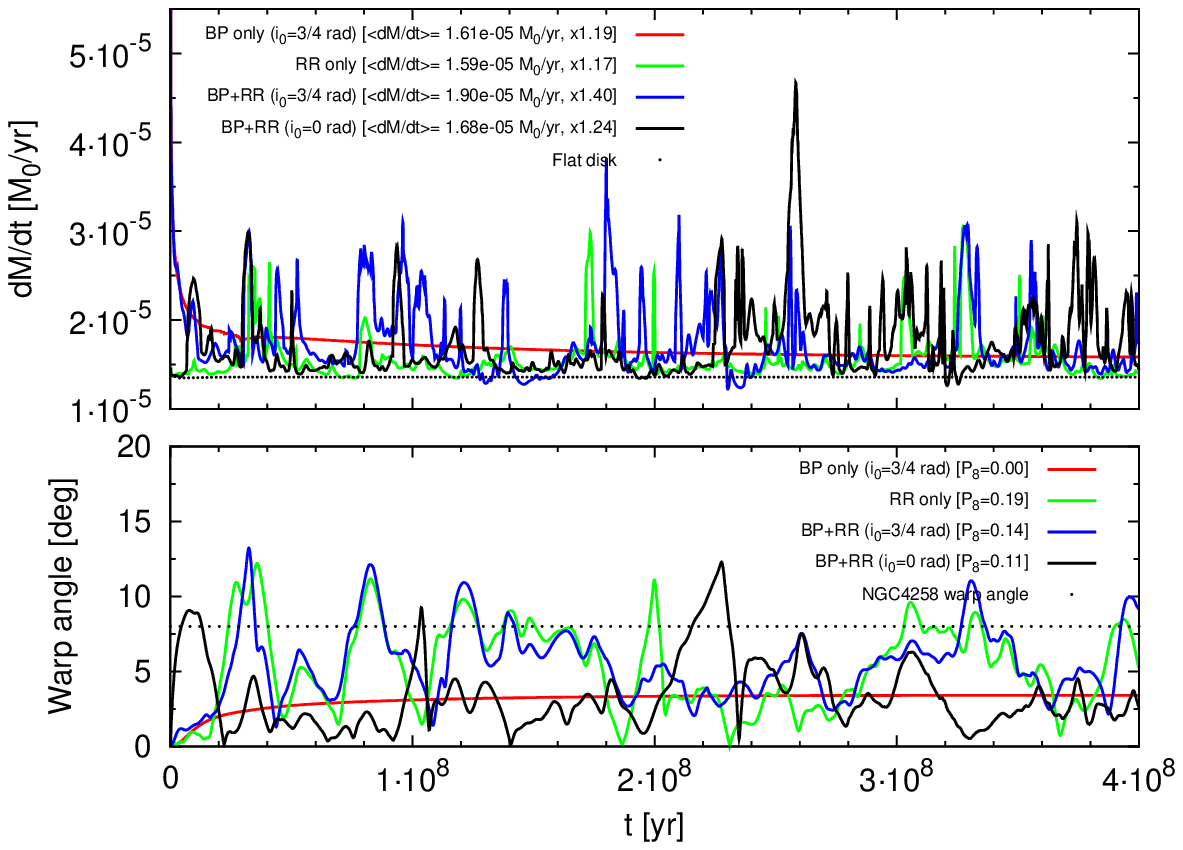}\tabularnewline
\end{tabular}
\par\end{raggedright}

\caption{\label{f:wmaser} Simulated disk evolution with $\alpha_{1}=0.25$,
$\chi=0$ or $1$ and with Kramer's opacity law, for the cases of
BP warping only ($N_{\star}=0$), RR warping only ($\chi=0$) and
both BP and RR warping (all models have the same RR realization and
initial MBH spin $\mathbf{J}_{\bullet}=\hat{z}$ and $i_{0}=0$ or
$3/4$ rad. They were evolved over $10^{9}$ yr, of which the initial
$4\times10^{8}$ yr are shown). Top: The mass accretion rate. The
mass loss rate from the equivalent steady state flat disk, $\dot{M}=1.4\times10^{-5}\,\Mo\,\mathrm{yr^{-1}}$,
is also shown for reference. The mean mass loss rates, and their enhancement
over that of a steady state flat disk are indicated in the plot labels.
Bottom: The warp angle across the maser zone. The best fit warp angle
of $\omega=8^{\circ}$ for NGC\,4258 \citep{her+05} is shown for
reference. The fraction of time the models spend with a warp as large
as that observed, $P_{8}$, are indicated in the plot labels. }
\end{figure*}

Figure (\ref{f:wmaser}) shows that RR-induced warping can easily
reproduce the observed $8^{\circ}$ warp angle across the maser region
of NGC\,4258 \citep{her+05}, without requiring any particular initial
conditions. The three scenarios explored here, RR with maximal ($\chi=1$)
frame-dragging and a large ($i_{0}=3/4$ rad) initial tilt, or with
a small ($i_{0}=0$ rad) initial tilt, or without frame-dragging ($\chi=0$),
all exhibit a probability $P_{8}\sim O(1)$ to have $\omega\ge8^{\circ}$
(see probabilities quoted in Figure \ref{f:wmaser}). This should
be contrasted with the scenario where only frame dragging is operating
\citep{cap+07}. In the example shown here, the warp angle never exceeds
$5^{\circ}$. By careful choice of the initial parameters, it may
be possible to obtain a large enough warp angle, and satisfy the constraints
set by the direction of the radio jet with frame dragging only. However,
this requires considerable fine-tuning, and may require a very long-lived
disk for the warp to grow \citep{mar08}.

The warping probability increases with the steepness of the stellar
cusp, as more stars are available near the disk to torque it. For
example, $P_{8}=0.02,$ $0.11$ and $0.29$ for $i_{0}=0$ rad $\chi=1$
in a sequence of models with $\gamma=3/2,$ $7/4$ and $2$, respectively.

We find that for the $i_{0}=0$ rad models of NGC\,4258, the rms
RR-induced misalignment between the MBH spin axis and the disk's total
angular momentum is $\mathrm{rms}(i)=15^{\circ}$ after $10^{8}$
yr for the case $\chi=1$, and as large as $\mathrm{rms}(i)=44^{\circ}$
for the $\chi=0$ case. This is an example of the stabilizing effect
of the BP torques, which introduce a preferred plane for the disk,
and suppress warps. It is interesting to note that the observed maser
disk in NGC\,4258 is tilted by $\sim30^{\circ}$ to the present jet
axis, as identified by the North and South hotpots \citep{cec+00,wil+01}.
Assuming that the jet is aligned with the MBH spin axis, then such
a tilt is consistent with RR torquing of a disk (see Figure \ref{f:snaps}d),
possibly around a sub-maximal spinning MBH.

\begin{figure}
\noindent \centering{}\includegraphics[clip,width=1\columnwidth]{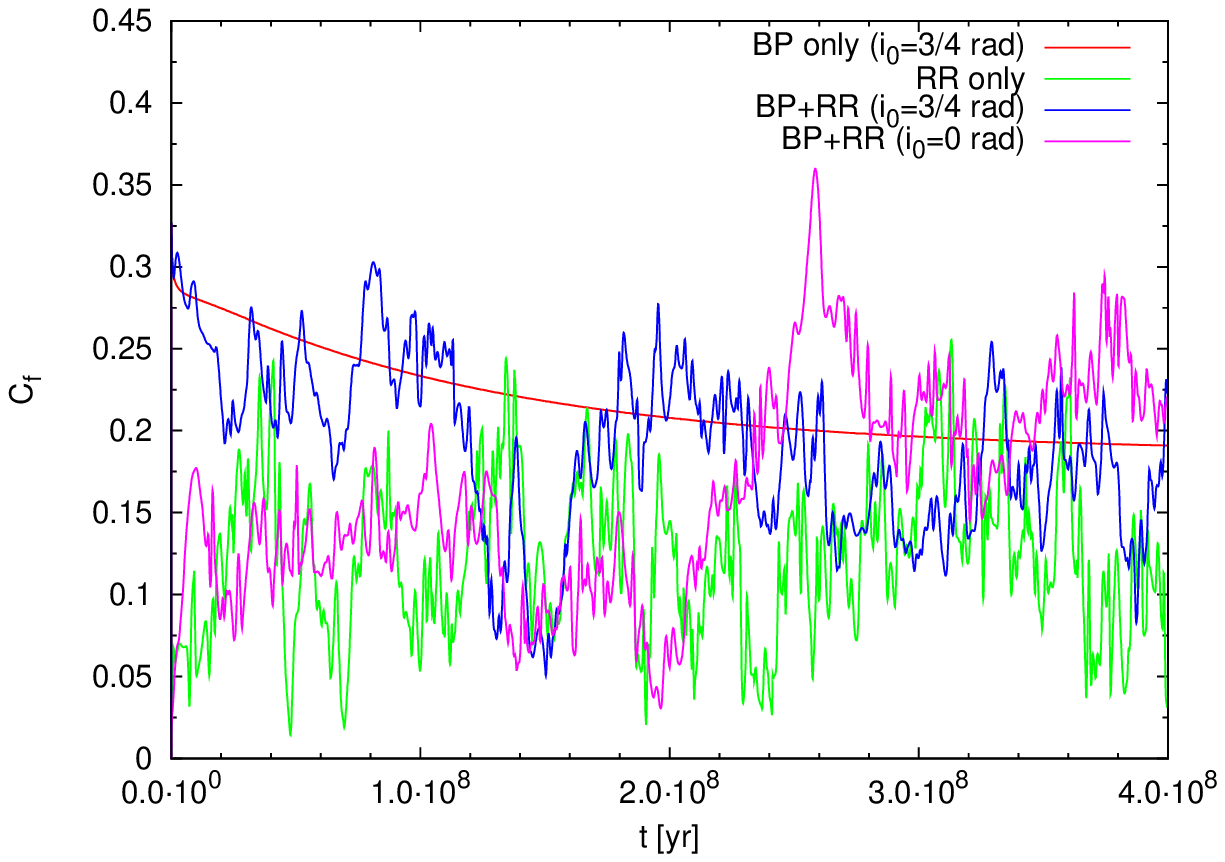}\caption{\label{f:Cf}The evolution of the maser disk's covering factor due
to BP and RR for the models shown in Figure (\ref{f:wmaser}).}
\end{figure}

The warping of the disk exposes it to ionizing radiation from its
innermost parts, which may play a central role in determining the
physical conditions in the outer regions of the disk. The fraction
of the central luminosity that falls on the disk (assuming isotropic
emission) is expressed by the disk's covering fraction, $C_{f}=\int_{0}^{2\pi}\mathrm{d}\phi\left|\cos\theta_{\max}(\phi)-\cos\theta_{\min}(\phi)\right|/4\pi$,
where $\theta_{\min}$ and $\theta_{\max}$ are the minimal and maximal
inclinations above and below the disk's mean plane along $R$ in azimuthal
direction $\phi$. Figure (\ref{f:Cf}) shows that RR-induced warping
gives the disk a minimal varying covering factor of $C_{f}\sim0.1-0.3$,
on top of any contribution from a pre-existing large scale warp (e.g
by the BP effect).

\subsection{Mass accretion rate}

\label{ss:Mdot}

\begin{figure*}[t]
\noindent \begin{centering}
\begin{tabular}{cc}
\includegraphics[width=0.5\textwidth]{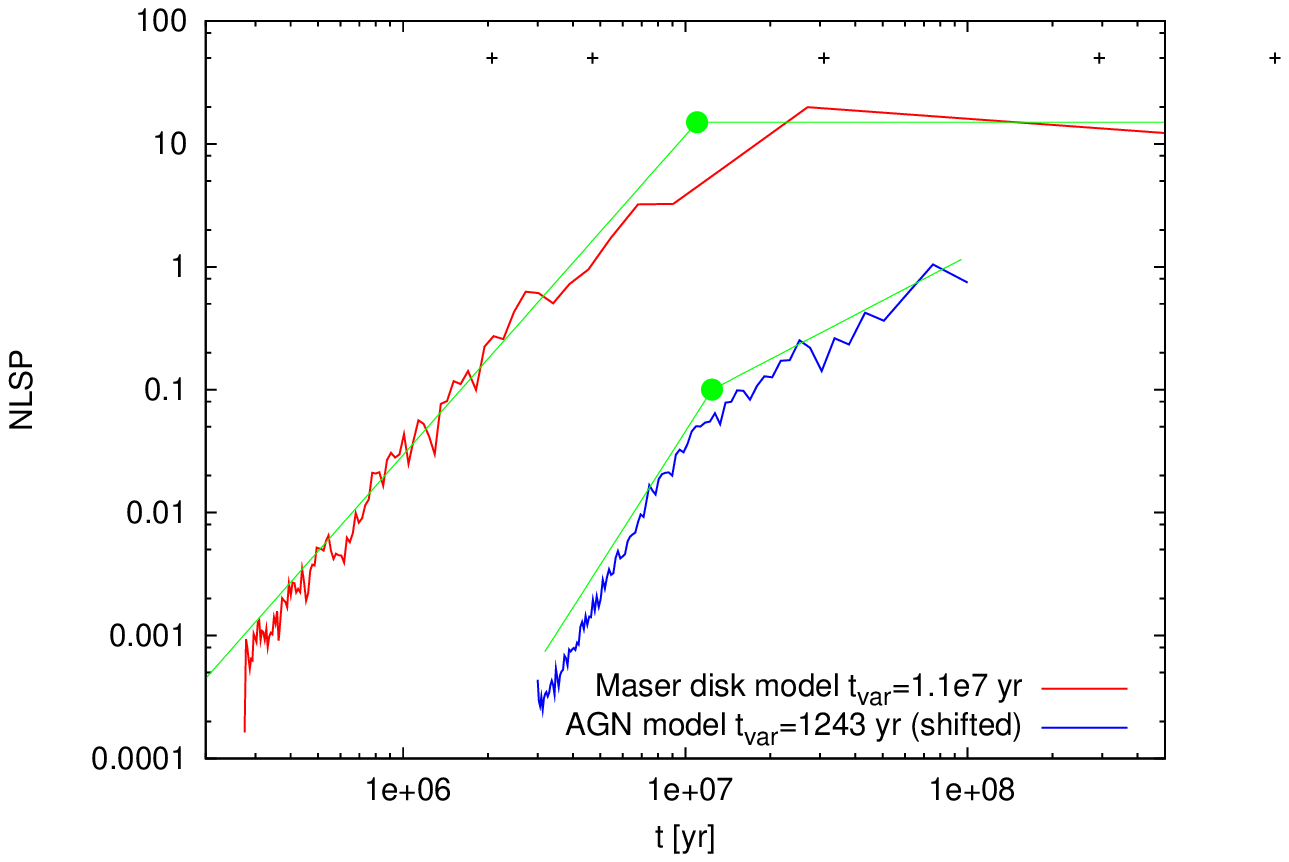} & \includegraphics[width=0.5\textwidth]{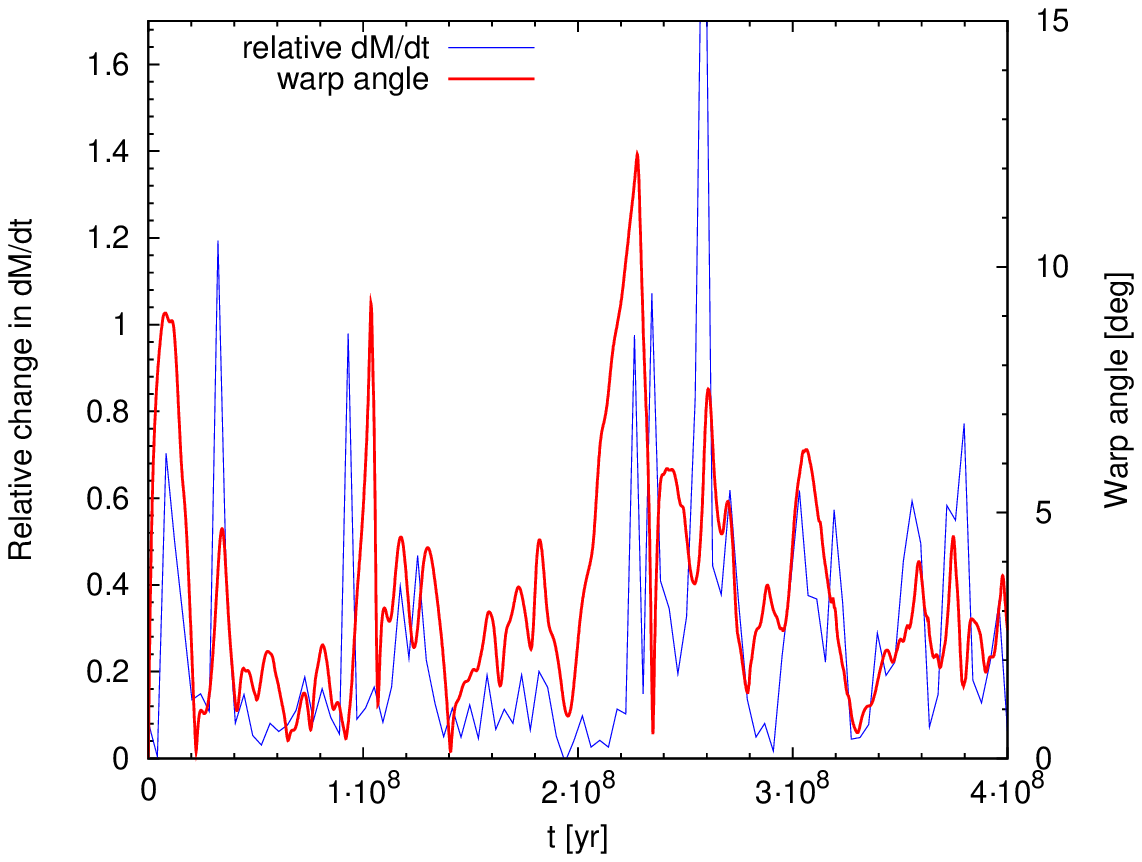}\tabularnewline
\end{tabular}
\par\end{centering}

\caption{\label{f:Mdot}Left: The smoothed Normalized Lomb-Scargle Periodograms
(power spectra) of $\dot{M}(t)$, as function of inverse frequency,
for the NGC~4258 maser disk model and for the AGN model discussed
in Section \ref{ss:Jbh}. The coherence times of the star shells that
are used to simulate the RR torques (Section \ref{ss:RRcusp}) for
the maser disk model are marked by crosses. The periodograms were
smoothed to highlight the power-law break at $t_{\mathrm{var}}$ (green
line and circle), which indicates that most of the power is at $t\gtrsim t_{\mathrm{var}}=(\Mbh/M_{d})P(R_{d})$
(The AGN model is displayed shifted on the logarithmic scale for comparison
with the maser disk model). Right: The correlation between the variability
on the large (maser region) scale and the relative mass accretion
rate (normalized to the steady state case) in a model of NGC~4258
(BP and RR, $i_{0}=0$ rad and $\chi=1$). The mass accretion rate
curve was smoothed by cubic spline, to filter out the short time-scale
variability (cf Figure \ref{f:wmaser}) and highlight the correlation
between warping and mass accretion rate on longer time scales.}
\end{figure*}

As noted by \citet{lod+06}, warping increases the mass accretion
rate over that in a flat disk because of the additional dissipation
due to the vertical shear. We confirm here that the warps are associated
with mass accretion fluctuations of up to factors of $3$--$4$ over
that in a flat disk (Figure \ref{f:wmaser}), resulting in an overall
increase by a factor of up to $f_{\dot{M}}\sim1.4$ in the average
mass accretion rate. The power spectrum of the accretion rate fluctuations
in the maser model is very broad (Figure \ref{f:Mdot}), and can be
roughly approximated by a broken power-law, which flattens beyond
$ $$\gtrsim10^{7}$ yr. This timescale corresponds to the RR coherence
time in the outer half of the disk, $ $$r\lesssim R_{d}=0.16$ pc
to $r\sim R_{2}=0.27$ pc. A rough estimate of this variability timescale
can be obtained by evaluating the coherence time at the disk's mass-weighted
mean radius, since that is where the gravitational coupling with the
stars is maximal. That length-scale is typically in the regime where
the back-reaction time is shorter than the self-quenching time (Eq.\ref{e:treact}),
and so 
\begin{equation}
t_{\mathrm{var}}\sim t_{\mathrm{react}}(R_{d})\sim(\Mbh/M_{d})P(R_{d})\simeq1.1\times10^{7}\,.\label{e:tvar}
\end{equation}
A similar correspondence is seen between the variability timescale
of the AGN model (Section \ref{ss:Jbh}), $t_{\mathrm{var}}\simeq1240$,
and the power-law break in its power spectrum (Figure \ref{f:Mdot}).
Thus, $t_{\mathrm{var}}$ can be interpreted as the shortest timescale
for which there is substantial variability power.

\subsection{MBH spin evolution}

\label{ss:Jbh}

\begin{figure}
\noindent \begin{centering}
\includegraphics[clip,width=1\columnwidth]{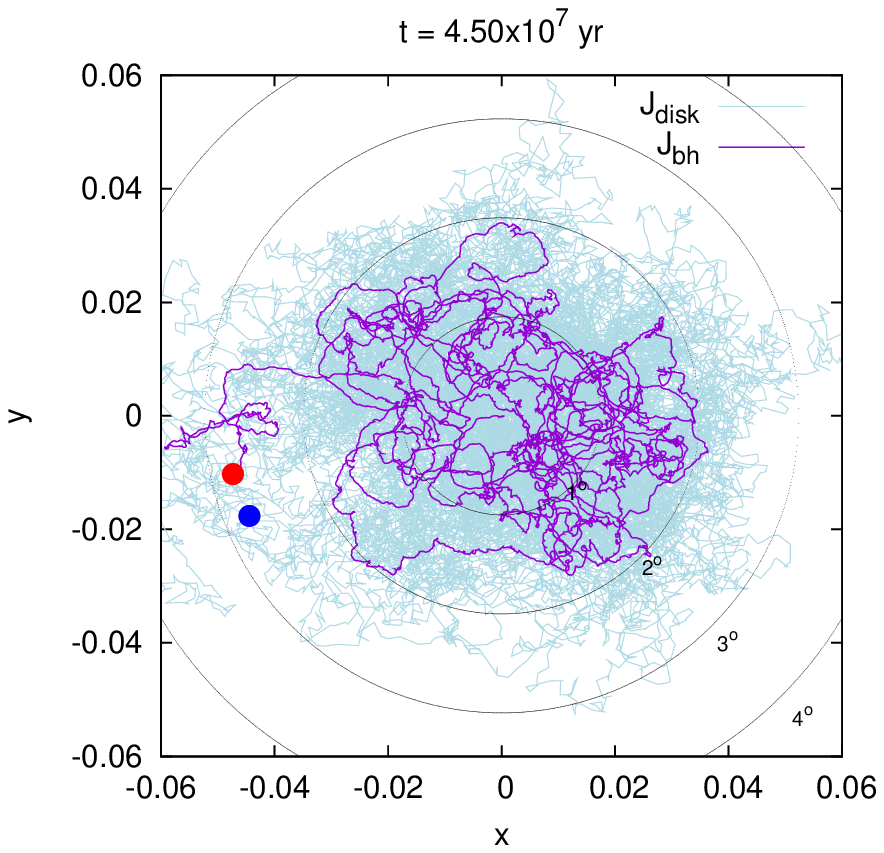} 
\par\end{centering}

\caption{\label{f:Jbh}The evolution of the MBH spin $\mathbf{J}_{\bullet}$,
and the disk's total angular momentum $\mathbf{J}_{\mathrm{disk}}$,
over an $e$-folding time (for $\eta=0.1$), shown projected on the
$(x,y)$ plane, under the influence of BP and RR together, for a low
mass AGN model with $\Mbh=4\times10^{6}\,\Mo$, $r_{h}=2.3$ pc, $\mu_{h}=2$,
$\gamma=2$, $\alpha_{1}=0.1$, $\chi=1$, Kramer's opacity law, normalized
with $[H/R]_{a}=0.002$, $\rho_{a}=2.4\times10^{-12}$, $T_{a}=1000\,\mathrm{K}$
and $\kappa_{a}=10\,\mathrm{cm^{2}\, g^{-1}}$ at $R_{a}=0.1$ pc,
and the initial conditions $\mathbf{J}_{\bullet}=\hat{z}$ and $i=0$
rad. These parameters result in $M_{d}=1.6\times10^{4}\,\Mo$, $R_{d}=2.2\times10^{-3}$
pc, $t_{\mathrm{var}}\simeq1240$ yr, $\dot{M}=0.013\,\Mo\,\mathrm{yr^{-1}=1.4\eta}\dot{M}_{E}$,
$R_{\mathrm{BP}}=3.6\times10^{-4}$ pc, and $t_{\parallel}=4.1\times10^{5}\,\mathrm{yr}$.
The blue and red circles denote the projection of $\mathbf{J}_{\bullet}$
and $\mathbf{J}_{\mathrm{disk}}$, respectively, at the end of the
simulation.}
\end{figure}

The enhanced mass accretion rate translates directly to a faster rate
of change in the magnitude of the the MBH spin, $\dot{J}_{\bullet}\simeq\dot{M}\sqrt{G\Mbh R_{\alpha}}$,
as the MBH spin vector remains aligned with the inner regions of the
disk due to the BP torques. The BP torques also couple the fluctuating
RR torques on the disk to the MBH spin orientation (Figure \ref{f:Jbh},
\ref{e:dJbhdt}). For the effect to be significant, the disk must
be massive enough to carry a substantial amount of angular momentum
(Eqs. \ref{e:Tbp}, \ref{e:dJbhdt}), but not so massive as to remain
effectively immobile against the RR torques. 

We find that the response of the MBH spin in NGC~4258 to the BP torques
is very small, because of the very low mass of the disk. However,
the effect can be larger in systems with a more massive accretion
disk. Figure (\ref{f:Jbh}) shows as an example the evolution of the
MBH spin axis relative to its initial direction (we omit the much
smaller spin evolution due to accretion), and that of the total angular
momentum of the disk, for a low-mass luminous AGN model that corresponds
to a Seyfert galaxy ($L\sim7\times10^{43}\,\mathrm{erg\, s^{-1}}$
for a radiative efficiency of $\eta=0.1$). The angular momenta of
the MBH and the disk both execute a random walk around their initial
orientation. The amplitude of the disk angular momentum jitter is
larger than that of the MBH spin, because the RR coherence time, $t_{\mathrm{react}}=620$
yr at $R_{d}=2.2\times10^{-3}$ pc is much shorter than the BP timescale
$t_{\parallel}=4.1\times10^{5}\,\mathrm{yr}$, and so the MBH spin
cannot {}``catch up'' with the faster changes in the direction of
the stellar torques. We find that for this specific model, the MBH
spin scatters across $\sim5^{\circ}$ over an $e$-folding time ($4.5\times10^{7}$
years). Another consequence of the short coherence time (compared
to $t_{\mathrm{warp}}(R_{d})\simeq3.5\times10^{4}$ yr) is that the
RR-induced warps in this disk model have angles of only $\sim1^{\circ}$.
Their effect on the mass accretion rate is correspondingly small,
only a $\sim2\%$ increase over that in a flat disk.

\section{Discussion and summary}

\label{s:discussion}

\subsection{Discussion}

A circumnuclear accretion disk does not exist in isolation around
a MBH. Rather, it shares the volume with the high-density nuclear
cluster that is predicted to form there while the nucleus evolves.
The cluster is expected to be on average spatially isotropic near
the MBH, but the Poisson fluctuations in the distribution of orbital
inclinations lead to a slowly varying residual force that exerts coherent
torques on the disk (RR). We argue (Section \ref{ss:OOM}) that a
thin Keplerian disk is primarily torqued by these purely gravitational
interactions with the stars, rather than by the hydrodynamical ones
that occur when stars plunge through the disk. The RR torques warp
the disk and can lead to order unity variability in the disk geometry
(Figs. \ref{f:snaps}, \ref{f:wmaser}), its mass accretion rate (Figs.
\ref{f:wmaser}, \ref{f:Mdot}) and covering factor (Figure \ref{f:Cf}).
The strong coupling between the stellar potential fluctuations and
the disk via RR is then further extended to the MBH spin via the frame-dragging
BP effect, and this allows angular momentum to be transferred from
the nuclear cluster to the MBH. The combined effects of the perpendicular
RR and BP torques excite a jitter in the MBH spin direction (Figure
\ref{f:Jbh}). In addition, the increased mass accretion rate due
to the RR-induced warping, together with the disk / MBH spin alignment
due to the BP effect, lead to an increased growth rate of the MBH
spin magnitude. We conclude that gravitational interactions between
the stars and the disk excite a substantial level of irreducible variability
in the disk properties---stationary accretion in a circumnuclear disk
is merely an idealization.

The large-scale warping of the disk (for NGC~4258, this coincides
with the maser region warp) reflects the RR coherence timescale at
the mass-weighted mean disk radius, $t_{\mathrm{var}}\sim(\Mbh/M_{d})P(R_{d})$
yr ($\sim10^{7}$ yr for NGC~4258). Generally, the effect of RR on
the disk and the MBH will be substantial when the disk is long-lived,
$t_{\mathrm{disk}}>t_{\mathrm{var}}$. There is a large spread in
the estimates of AGN and QSO lifespans in the luminous phase, $\mbox{\textrm{few}\ensuremath{\times}}10^{6}$--$10^{8}$
yr (e.g. \citealt{gra+04,por+04b}), with some recent analyses suggesting
an even longer lifespan of $\lesssim10^{9}$ yr \citep{gil+09,ros+09,sch+09}.
For long-lived disk systems, the cumulative effects of RR-torquing
can be large. 

One implication of the factor $f_{\dot{M}}\sim1.2-1.4$ increase in
the mean mass accretion rate over that in a flat stationary disk (Section
\ref{ss:Mdot}), is that an MBH fed by an RR-torqued disk, whose mass
accretion rate is raised to the Eddington limit by warping, will be
$\exp[(f_{\dot{M}}-1)(t_{\mathrm{disk}}/t_{E})]$ more massive than
one that is fed by an otherwise identical flat disk, which is accreting
at only $1/f_{\dot{M}}$ of the Eddington limit. For example, RR-torquing
can accelerate MBH growth over a time $t_{\mathrm{disk}}=10t_{E}=4.5\times10^{8}\,\mathrm{yr}$
(assuming $\eta=0.1$) by a factor of up to $\sim50$ (for $f_{\dot{M}}=1.4$).
Alternatively, if radiation pressure prevents the increased RR-induced
mass flow from accreting on the MBH, the accumulating excess mass
may trigger outflows, or disk fragmentation and star formation.

Another cumulative effect of RR-torquing of the disk is the displacement
of the MBH spin from its initial orientation due to the BP-induced
random jitter. It is interesting to note that the few degrees amplitude
of the displacement is still consistent with the typical observed
$\sim5^{\circ}$ opening angle of AGN radio jets on large scales \citep{opp+94},
assuming the jet direction reflects the spin direction.

A direct geometric consequence of RR-induced warping is the substantial
time-averaged covering factor the disk acquires relative to the central
continuum source. This allows the disk to intercept the central radiation
and be heated and ionized by it. Such X-ray irradiation may in fact
be essential for raising the gas temperature to the range necessary
for maser emission \citep{neu+95}, for the production of the observed
emission lines in AGN disks in general \citep{col87}, for driving
winds and outflows from disks \citep{pro+04}, or for explaining Narrow
Line Seyfert 1 galaxies and ultra-soft AGN \citep{puc+02}. 

The magnitude of the effects of the stars on the disk depend on the
central concentration of the stellar cluster, and is therefore sensitive
to the degree of mass segregation in the system and to the fraction
of massive compact remnants in the population. The more centrally
concentrated the stellar density profile, and the more massive the
stars (for RR, the relevant quantity is $\left\langle \Ms^{2}\right\rangle /\left\langle \Ms\right\rangle $,
\citealt{rau+96}), the larger are the effects on the disk. It is
worth noting that strong mass segregation with a power-law cusp profiles
of $\gamma\gtrsim2$ is expected to occur in nuclei with old stellar
populations \citep{ale+09,kes+09,pre+10}, where the inner parts of
the cusp are dominated by stellar black holes of mass $O(10\,\Mo)$.
It is thus plausible that a substantial fraction of galactic nuclei
have conditions that are conducive to RR-torquing of an accretion
disk.

In this study we presented numerical results for galactic nuclei with
MBH masses in the range $\sim4\times10^{6}\,\Mo$ (with a massive
disk close to $\dot{M}_{E}$) to $\sim4\times10^{7}\,\Mo$ (with a
low-mass, low-accretion rate disk). In order to scale the results
to other systems, it is necessary to specify the scaling of the disk
properties with MBH mass. We defer this to future work, and present
here only a simple preliminary analysis that suggests that the effects
are generally more important in lower-mass MBHs. Assuming that all
AGN disks extend up to some universal large multiple of $r_{g}$ (e.g.
$\sim2000r_{g}$ for $\alpha$-disks limited by gravitational instability,
\citealt{goo03}), and assuming they all have the same aspect ratio,
then $R_{d}\propto\Mbh$ and $M_{d}/\Mbh\simeq H/R$, so that $M_{d}\propto\Mbh$.
The angular momentum in the disk therefore scales as $J_{d}\propto M_{d}\sqrt{\Mbh R_{d}}\propto\Mbh^{2}$.
The $\Mbh/\sigma$ relation indicates that the MBH radius of influence
scales as $r_{h}\propto\Mbh^{1/2}$ (assuming $\Mbh\propto\sigma^{4}$),
so that the number of stars in the MBH cusp inside $R_{d}$ scales
as $\Ns(R_{d})\propto\Mbh(R_{d}/r_{h})^{3-\gamma}\propto\Mbh^{(5-\gamma)/2}$.
The residual angular momentum in the stars that are efficiently coupled
to the disk then scales as $J_{N}\propto\sqrt{\Ns(R_{d})}\sqrt{\Mbh R_{d}}\propto\Mbh^{(9-\gamma)/4}$,
and it then follows that $J_{d}/J_{N}\propto\Mbh^{(\gamma-1)/4}$.
The ratio of the disk and stellar angular momenta is a slowly rising
function of $\Mbh$ for $\gamma>1$. This implies that, all other
parameters being equal, it should become progressively more difficult
for the stars to torque the disk as the MBH mass increases. In addition,
as argued in Section (\ref{ss:OOM}), the larger $\Mbh$, the more
the disk responds to the RR torques as a rigid body, rather than by
growing warps.

\subsection{Caveats}

The calculations and conclusions presented here are limited by various
assumptions and approximations. They are listed here briefly. Some
of these issues will be addressed in future work.

\paragraph{Limitations of the galactic nucleus model}

The RR torques by the stellar cluster are approximated by the gravitational
field of a small number of thin rings (Section \ref{ss:RRcusp}),
which provides only a rough approximation of the true power spectrum
of the stellar perturbations. The back-reaction of the disk on the
stars is approximated simplistically by limiting the RR coherence
time so it does not exceed the back-reaction time. A more realistic
fluctuation spectrum will allow, among other issues, to better model
the very short timescale fluctuations, and explore whether these have
any relation to the observed optical/UV variability of AGN. Our conclusions
on mass accretion rates and MBH spin evolution are generalizations
based on extrapolating simulations that were constructed explicitly
to represent the disk and cluster of NGC\,4258. They should be scaled
and tested for a wider range of MBHs, disks and nuclear clusters .

\paragraph{Limitations of the disk model}

The physics underlying accretion disks are still not well understood.
In particular the nature of the azimuthal and vertical viscosities,
and the relation between them, are quite uncertain \citep{ogi99,lod+07,lod+10}.
The disk is not modeled self-consistently. The thermal structure of
the disk is based for simplicity on a Kramer opacity law with a free
normalization constant, which is clearly non-physical, as indicated
by the very high opacity that is required to produce the masing conditions.
External X-ray heating, which is likely important, is neglected. MBH
spin evolution due to mass accretion can be relevant on longer timescales,
but is not taken into account, and the artificial BP torque suppression
on small scales that is needed to stabilize the results numerically,
may both lead to an underestimation of the effect of disk warping
on the MBH spin evolution. Finally, the disk is approximated as Newtonian
and Keplerian, even near the ISCO, which is held fixed at its Schwarzschild
value.

\paragraph{Limitations of the numerical scheme}

The numerical scheme used here \citep{pap+83,pri92} can not describe
azimuthal modes and as implemented, is limited to moderate warp angles
(although viscosities for arbitrary large angles can be calculated
numerically, \citealt{ogi99}). In particular, disk flipping from
co-rotation to counter-rotation, or disk destruction by large deformations,
cannot be modeled reliably.

\subsection{Summary}

We analyzed and simulated numerically the evolution of a thin accretion
disk around a MBH that is surrounded by a stellar cluster. We took
into account the disk's internal viscous torques, the frame-dragging
torques of a spinning MBH and the stellar orbit-averaged gravitational
torques. We show that the evolution of the MBH mass accretion rate,
the MBH spin growth rate, and the covering fraction of the disk relative
to the central ionizing continuum source, are all strongly coupled
to the stochastic fluctuations of the stellar potential via the warps
that the stellar torques excite in the disk. These lead to fluctuations
by factors of up to a few in these quantities over a wide range of
timescales, with most of the power on timescales $\gtrsim(\Mbh/M_{d})P(R_{d})$.
The response of the disk is stronger the lighter it is and the more
centrally concentrated the stellar cusp. We demonstrated these effects
by simulating the evolution of the maser disk in NGC~4258, and show
that its observed $O(10^{\circ})$ warp can be driven by the stellar
torques. We also show that the frame-dragging of a massive AGN disk
couples the stochastic stellar torques to the MBH spin and can excite
a jitter of a few degrees in its direction relative to that of the
disk's outer regions.

\acknowledgements{We thank J. Cuadra, J. Granot, K. G\"ultekin J.-P. Lasota, A. Levinson,
G. Lodato, C. Nixon and F. Pedes for helpful discussions and comments.
T.A. acknowledges support by ERC Starting Grant 202996 and DIP-BMBF
grant 71-0460-0101.}

\appendix{}

\section{A. The discretized disk evolution equation}

\label{a:discrete}

The integration of the evolution equation generally follows the scheme
of \citet{pri92}, but some details of the implementation are different.
It is presented here briefly for completeness. 

All quantities are expressed in a system of units where $G=c=M_{\bullet}=1$.
Denoting by $\Delta z$ the equal spacing of the logarithmic grid,
the grid points are at $R_{i}=R_{1}e^{(i-1)\Delta z}$ for $i=0,\ldots,N+1$,
where points $R_{0}$, $R_{N+1}$ represent the boundaries and $R_{1}=6$
is at the ISCO. The angular momenta densities at time $t_{j}$, $\mathbf{L}_{i}^{j}$
($i=1,\ldots,N$), are advanced in time to $t_{j+1}=t_{j}+\Delta t_{j}$
by

\begin{eqnarray}
\mathbf{L}_{i}^{j+1} & = & \mathbf{L}_{i}^{j}+\frac{\Delta t_{j}}{(\Delta z)^{2}R_{i}^{2}}\nonumber \\
 &  & \times\left\{ 3\left[\left(\nu_{1,i+1}^{j}L_{i+1}^{j}-\nu_{1,i}^{j}L_{i}^{j}\right)\boldsymbol{\bar{\ell}}_{i,i+1}^{j}-\left(\nu_{1,i}^{j}L_{i-1}^{j}-\nu_{1,i-1}^{j}L_{i-1}^{j}\right)\boldsymbol{\bar{\ell}}_{i-1,i}^{j}\right]\right.\nonumber \\
 &  & +\frac{1}{2}\left[\bar{\nu}_{2,i,i+1}^{j}\bar{L}_{i,i+1}^{j}(\boldsymbol{\ell}_{i+1}^{j}-\boldsymbol{\ell}_{i}^{j})-\bar{\nu}_{2,i-1,i}^{j}\bar{L}_{i-1,i}^{j}(\boldsymbol{\ell}_{i}^{j}-\boldsymbol{\ell}_{i-1}^{j})\right]\nonumber \\
 &  & -\Delta z\left[V_{\mathrm{adv,}k+1}^{j}R_{k+1}\mathbf{L}_{k+1}^{j}-V_{\mathrm{adv,}k}^{j}R_{k}\mathbf{L}_{k}^{j}\right]\\
 &  & \left.+\left[\bar{\nu}_{3,i,i+1}^{j}\bar{\mathbf{L}}_{i,i+1}^{j}\times(\boldsymbol{\ell}_{i+1}^{j}-\boldsymbol{\ell}_{i}^{j})-\bar{\nu}_{3,i-1,i}^{j}\bar{\mathbf{L}}_{i-1,i}^{j}\times(\boldsymbol{\ell}_{i}^{j}-\boldsymbol{\ell}_{i-1}^{j})\right]\right\} \nonumber \\
 &  & +\Delta t_{j}\mathbf{T}_{\mathrm{ext},i}^{j}\,,
\end{eqnarray}
where $\bar{\mathbf{L}}_{i,i+1}^{j}=(\mathbf{L}_{i}^{j}+\mathbf{L}_{i+1}^{j})/2$,
$\boldsymbol{\bar{\ell}}_{i,i+1}^{j}=\bar{\mathbf{L}}_{i,i+1}^{j}/\left|\bar{\mathbf{L}}_{i,i+1}^{j}\right|$,
and $\bar{\nu}_{n,i,i+1}^{j}=(\nu_{n,i}^{j}+\nu_{n,i+1}^{j})/2$ for
$n=2,3$. The index $k$ is defined as $k=i-1$ when the advective
velocity $V_{\mathrm{adv},i}^{j}>0$, and $k=i$ when $V_{\mathrm{adv},i}^{j}<0$,
where 
\begin{equation}
V_{\mathrm{adv},i}^{j}=\frac{1}{R_{i}}\left[\frac{3}{2}\nu_{1,i}-\nu_{2,i}\left|\frac{\boldsymbol{\ell}_{i+1}-\boldsymbol{\ell}_{i-1}}{2\Delta z}\right|^{2}\right]\,,\label{e:Vadv}
\end{equation}
for $i=1,\ldots N$. For $i=0$, $i-1$ in Eq. (\ref{e:Vadv}) is
substituted by $0$, and for $i=N+1$, $i+1$ is substituted by $N+1$.
The viscosities (and all other thermodynamic properties) are then
updated via Eq. (\ref{e:TcR}).

The inner boundary conditions, $\mathbf{L}(R_{1})=0,$ $\partial\boldsymbol{\ell}/\partial R|_{R_{1}}=0$
are enforced by setting $ $$\mathbf{L}_{0}^{j}=0$ (this causes a
slight deviation from the analytic solution, since $\mathbf{L}_{1}^{j}\rightarrow0$,
but is not zero identically), and $\boldsymbol{\ell}_{0}^{j}=\boldsymbol{\ell^{j}}_{1}$.
The outer boundary condition, $\partial(\nu_{1}\mathbf{L})/\partial R=0$
is enforced by setting $\mathbf{L}_{N+1}^{j}=(\nu_{1,N}^{j}/\nu_{1,N+1}^{j})\mathbf{L}_{N}^{j}$.
The mass-loss from the inner edge of the disk is balanced (in the
statistical sense) by adding a source term at the outer radius, $\mathbf{L}_{N}^{j}\rightarrow\mathbf{L}_{N}^{j}+\left(\Delta L_{N}\right)^{j}\boldsymbol{\ell}_{0}$
(see Section \ref{ss:ICBC} and Eq. \ref{e:dLN}).

The time-step size is adjusted every time-step to 
\begin{equation}
\Delta t_{j}=\min(\Delta t_{\nu_{1}}^{j},\Delta t_{\nu_{2}}^{j},\Delta t_{\nu_{3}}^{j},\Delta t_{\mathrm{BP}}^{j},\Delta t_{\mathrm{RR}}^{j})/2\,,
\end{equation}
where

\begin{equation}
\begin{array}{ll}
\Delta t_{\nu_{n}}^{j}=\min_{i}(\Delta R_{i})^{2}/\bar{\nu}_{n,i,i+1}^{j}\, & \mathrm{for\,}n=1,2,3\,,\\
\Delta t_{\mathrm{BP}}^{j}=\min_{i}L_{i}^{j}/|\mathbf{T}_{\mathrm{BP},i}^{j}|\,, & \Delta t_{\mathrm{RR}}^{j}=\min_{i}(\Ms/\Mbh)\left/\sqrt{N_{\star}(<R_{i})}R_{i}^{3/2}\right.\,,
\end{array}
\end{equation}
and where $\bar{\nu}_{1,i,i+1}$ is defined similarly to $\bar{\nu}_{2,i,i+1}$.

Angular momentum conservation is monitored by evaluating the change
in the total angular momentum of the disk between times $t_{j-1}$
and $t_{j}$ in two different ways. One, which corresponds to the
divergence term in the continuity equation, is $\Delta\mathbf{J}_{\mathrm{edge}}^{j}(\mathbf{L}_{0}^{j},\mathbf{L}_{1}^{j},\mathbf{L}_{N}^{j},\mathbf{L}_{N+1}^{j},\mathbf{L}_{0}^{j-1},\mathbf{L}_{1}^{j-1},\mathbf{L}_{N}^{j-1},\mathbf{L}_{N+1}^{j-1})=-2\pi\Sigma_{i=0}^{N+1}R_{i}\Delta R_{i}(\mathbf{L}_{i}^{j}-\mathbf{L}_{i}^{j-1})$,
which can be written as a function of the disk edges only, since all
the inner grid points cancel out in the sum. The other, which corresponds
to the time derivative term in the continuity equation, is the change
of angular momentum in the bulk, $\Delta\mathbf{J}_{\mathrm{bulk}}^{j}=\mathbf{J}_{\mathrm{bulk}}^{j}-\mathbf{J}_{\mathrm{bulk}}^{j-1}$,
where $\mathbf{J}_{\mathrm{bulk}}^{j}=2\pi\Sigma_{i=1}^{N}R_{i}\Delta R_{i}\mathbf{L}_{i}^{j}$.
The fractional degree of non-conservation per time-step in the absence
of source terms (Section \ref{ss:validation}) is then $\delta J^{j}=\left.\left|\Delta\mathbf{J}_{\mathrm{bulk}}^{j}+\Delta\mathbf{J}_{\mathrm{edge}}^{j}\right|\right/J_{\mathrm{bulk}}^{j-1}$.
Analogous expressions are used to monitor mass conservation.

\section{B. Second order coefficients of the viscosity parameter expansion}

\label{a:B123}

The expansion of the dimensionless viscosity parameters (Eq. \ref{e:a123})
was derived by \citet{ogi99}. The second order coefficients are reproduced
here for convenience in the notation of this paper, as functions of
the in-plane shear viscosity parameter $\alpha_{1}$ and the adiabatic
index $\Gamma$ only ($\Gamma=7/5$ for diatomic gas). The dependence
on a bulk viscosity parameter, assumed here to be zero, is not included.

\begin{equation}
B_{1}=\frac{-2(1-17\alpha_{1}^{2}+21\alpha_{1}^{4})}{12\alpha_{1}(4+\alpha_{1}^{2})}\,,\,\,\, B_{2}=2\mathrm{Re}(Q_{23})\,,\,\,\, B_{3}=\mathrm{Im}(Q_{23})\,,
\end{equation}

\[
\]
 where 
\begin{equation}
Q_{23}=\frac{\tilde{a}+\tilde{b}\Gamma+\tilde{c}[\Gamma-(8/3)\alpha_{1}i]}{4\alpha_{1}[3-\Gamma+(8/3)\alpha_{1}i](-\alpha_{1}+2i)^{3}(\alpha_{1}+2i)}\,,
\end{equation}
 and
\begin{eqnarray}
\tilde{a} & = & (12+49\alpha_{1}^{2}-408\alpha_{1}^{4}+2\alpha_{1}^{6})+(115+109\alpha_{1}^{2}-41\alpha_{1}^{4}+4\alpha_{1}^{6})\alpha_{1}i\,,\nonumber \\
\tilde{b} & = & (18-87\alpha_{1}^{2}-24\alpha_{1}^{4})+(87-6\alpha_{1}^{4})\alpha_{1}i\,,\nonumber \\
\tilde{c} & = & (12-36\alpha_{1}^{2}+140\alpha_{1}^{4}+2\alpha_{1}^{6})+(25-11\alpha_{1}^{2}+21\alpha_{1}^{4})\alpha_{1}i\,.
\end{eqnarray}
$ $

\bibliographystyle{apj}

\end{document}